\documentclass[aps,prxquantum,reprint,amsmath,floatfix,amssymb,superscriptaddress]{revtex4-2}
\usepackage{graphicx}
\usepackage{dcolumn}
\usepackage{bm}
\usepackage{color}
\usepackage{hyperref}
\usepackage{float}
\usepackage{dsfont}
\usepackage{mathtools}
\hypersetup{colorlinks,allcolors=blue}
\usepackage{orcidlink}
\usepackage{physics}
\newcommand{\be}{\begin{equation}}
\newcommand{\ee}{\end{equation}}
\newcommand{\bea}{\begin{eqnarray}}
\newcommand{\eea}{\end{eqnarray}}
\newcommand{\ba}[1]{\begin{array}{#1}}
\newcommand{\ea}{\end{array}}
\usepackage{booktabs}

\begin{document}
\title{Selective Preparation of Collective States in Coupled Quantum Emitters Using the SUPER Excitation Scheme}
\author{Johannes Kerber\,\orcidlink{0009-0002-1957-8008}}
\thanks{Contact author: Johannes.Kerber@uibk.ac.at}
\affiliation{Institut für Theoretische Physik, Universität Innsbruck, Technikerstraße 21a, A-6020 Innsbruck, Austria}
\author{Laurin Ostermann\,\orcidlink{0000-0001-8508-9785}}
\affiliation{Institut für Theoretische Physik, Universität Innsbruck, Technikerstraße 21a, A-6020 Innsbruck, Austria}
\author{Vikas Remesh\,\orcidlink{0000-0003-2029-0570}}
\affiliation{Institut für Experimentalphysik, Universität Innsbruck, Technikerstraße 25d, 6020 Innsbruck, Austria}
\author{Helmut Ritsch\,\orcidlink{0000-0001-7013-5208}}
\affiliation{Institut für Theoretische Physik, Universität Innsbruck, Technikerstraße 21a, A-6020 Innsbruck, Austria}
\author{Arpita Pal\,\orcidlink{0000-0003-1124-6818}}
\thanks{Contact author: Arpita.Pal@uibk.ac.at}
\affiliation{Institut für Theoretische Physik, Universität Innsbruck, Technikerstraße 21a, A-6020 Innsbruck, Austria}
\date{\today}

\begin{abstract}
The efficient preparation of collective eigenstates of subwavelength-spaced optical dipoles is a prerequisite for observing their signature radiative properties and for their applications in quantum information processing. We theoretically investigate the deterministic preparation of superradiant and subradiant states of two dipole-coupled two-level quantum emitters at deep-subwavelength separation using the Swing-UP of Quantum Emitter Population (SUPER) excitation scheme. Utilizing suitable pulse parameters for two red-detuned, time-overlapping Gaussian pulses, the SUPER scheme enables close-to-unity population inversion in the targeted collective eigenstates. Furthermore, a tunable optical phase in the SUPER scheme enables the simultaneous inversions in both pure super- and subradiant states with finite populations, thereby resulting in the preparation of hybrid collective states. These results are possible to realize with or without an optical cavity. Our approach to populating the collective eigenstates in a cavity environment paves the way for the efficient preparation of these states in the presence of environmental decoherence. Our scheme enables single-photon generation, which is measured using the second-order correlation function. We also discuss in detail possible experimental realizations, in particular using solid-state emitters and molecules.
\end{abstract}
\maketitle

\section{Introduction}
Collective light-matter interaction among subwavelength-spaced quantum emitters (QEs) is an intriguing research area~\cite{genes:prx:2022, Janne:PRA:2023} in nonlinear quantum optics. When QEs are positioned sufficiently closely, an individual QE begins to experience the electromagnetic presence of its neighbors through dipole-dipole interaction. This interaction gives rise to collective eigenstates with modified energy levels, entanglement, and distinct radiation properties. Consequently, the QEs' collective emission behavior exhibits superradiance, i.e., enhanced decay, and subradiance, i.e., suppressed decay, compared to the spontaneous emission of a single QE. Historically, following the seminal work of Dicke~\cite{Dicke:pr:1954} in 1954, significant research efforts were dedicated to investigating these collective eigenstates~\cite{haroche:1982, brewer:prl:1996, vahid:science:2002, Zoubi_2010, gauger:nc:2014, asenjo:prx:2017, zoller:prl:2019, An:science:2018, rui:nature:2020, rabl:prl:2025}, both theoretically and experimentally, spanning the domain from few~\cite{wallraff:nc:2014, Gerardot:sc:2022, kirchmair:np:2022, lounis:nc:2022, lodhal:science:2023, Hood:np:2024} to many QEs~\cite{Hommel:np:2007, kaiser:prl:2016, Ferioli:prx:2021, hotter:prr:2023, Fasser:OQ:24}. Successful preparation of specific collective eigenstates opens up exciting possibilities, for instance, harnessing the unique radiative properties~\cite{Thompson:nature:2012, Ruostekoski:prl:2016, zoller:prl:2019, holzinger:prl:2020, Mattiotti:njp:2021, gauger:prx:2023, gauger:prl:2025} and enabling applications in quantum information processing~\cite{jelena:np:2009, hammerer:rmp:2010},  storage~\cite{Ruostekoski:prl:2016, Ferioli:prx:2021}, entanglement~\cite{santos:pra:2022, santos:prl:2023}, quantum metrology~\cite{genes:prl:2013, Treutlein:rmp:2018}, and sensing~\cite{zoller:prl:2024}.

To observe the characteristic radiative features of these collective states~\cite{sven:nc:2016,bohr:nc:2024, aizpurua:prr:2024}, in particular when dealing with few-emitters, precise experimental control over various factors is typically required. These include tailored internal state selection~\cite{Orrit:jasp:1999}, positioning of the QEs~\cite{dalacu:nl:2020, Gerardot:sc:2022, lounis:nc:2022, lodhal:science:2023, Hood:np:2024}, preparing a controlled or near-dissipation-free environment~\cite{trotta:prl:2018, rotenberg:pra:2022}, and suitable spatial designs to obtain radiative coupling through guided optical modes~\cite{lodhal:rmp:2015, lodhal:science:2023}. It is also vital to implement suitable optical control to probe the collective states~\cite{Gerardot:sc:2022, lounis:nc:2022, lodhal:science:2023, Hood:np:2024} and control of optical phases~\cite {lodhal:science:2023,van2025resonant}. Recent experimental efforts to observe such characteristic radiation properties include studies involving self-assembled solid-state emitters~\cite{Gerardot:sc:2022}, solid-state emitter-photonic crystal waveguide systems~\cite{lodhal:science:2023}, molecule-electrode systems~\cite{lounis:nc:2022}, molecules embedded as defects in crystals~\cite{Hood:np:2024}, and nanophotonic cavity-QE systems~\cite{kim:nc:2025}. 

A key question in this research frontier is how to selectively prepare these collective eigenstates~\cite{Genes:sr:2015, chen:pra:2016, Ferioli:prx:2021, lounis:nc:2022, kim:nc:2025}, enabling direct access to their distinct radiative features. In this article, we, for the first time, present a novel theoretical solution by demonstrating that deterministic and efficient population of targeted collective states is indeed possible using the \emph{Swing-UP of Quantum Emitter population} (SUPER) excitation scheme. This fast and pulsed excitation approach effectively penetrates the system's electromagnetic environment, reasonably overcoming the influences of dissipation during the population inversion process. Specific collective eigenstates, in particular dark, bright, as well as hybrid collective states, can be efficiently prepared, as per requirement.

SUPER is an off-resonant, yet coherent, two-color pulsed excitation scheme, where a pair of red-detuned, ultrashort Gaussian laser pulses induces an efficient population transfer from an energetically lower to a higher state~\cite{doris:PRXQ:2021, karli:nl:2022, Super:arxiv:2023, boos:at:2024, joss:nl:2024}, primarily in solid-state emitters. Apart from the fact that it is off-resonant, which simplifies the pump laser rejection to a simple spectral filtering, it offers several benefits over conventional resonant excitation methods. For instance, it minimizes decoherence caused by phonons~\cite{doris:pssb:2022} at low temperatures, and has the potential to generate almost ideal single photons reducing possibilities of re-excitation even in the strong coupling regime as well as entangled photon pairs with high concurrence~\cite{Schumacher:PRR:2024} even at elevated temperatures~\cite{Bracht:OQ:23} with quantum dot-cavity systems. Recent studies on the fundamental understanding of the excitation mechanism~\cite{doris:PRB:2023, Vannucci:oe:2024, doris:prr:2025} in individual quantum dots offer promising results for intriguing photonic applications.

As a novel contribution to utilize the SUPER excitation scheme in the realm of collective-light effects, we explore the theoretical possibilities of utilizing the SUPER scheme for the selective preparation of collective eigenstates, in particular, superradiant and subradiant states. This is achieved in a deep-subwavelength arrangement of two two-level QEs even amidst environmental decoherence. Our proposed theoretical study is not platform-specific and relies on a generic two-level QE description. Considering that two decay channels (superradiant and subradiant) emerge with collective states, the SUPER scheme's novelty lies in its ability to access both the short-lived `bright' channel and the long-lived `dark' channel. Our scheme, importantly, allows for the possibility to populate states in both channels to produce hybrid collective states just by controlling the relative optical phases. We provide a detailed theoretical discussion on how so-called mixture weights influence the outcome of a particular excitation process. The efficient preparation of these collective eigenstates offers novel opportunities to observe their signature dissipative features, in particular with and without a cavity interface~\cite{cqed:1989}. Furthermore, a cavity interface would offer opportunities to analyze photon statistics using the second-order correlation function and reveal signatures of single-photon generation. For completeness, we discuss in detail some realistic experimental possibilities, considering both state-of-the-art and foreseeable developments, involving platforms of solid-state emitters and molecules, to realize our theoretical proposal.

\section{Theoretical Description}
\label{sec:TD}
\begin{figure}[t]
\centering
\includegraphics[width=\linewidth]{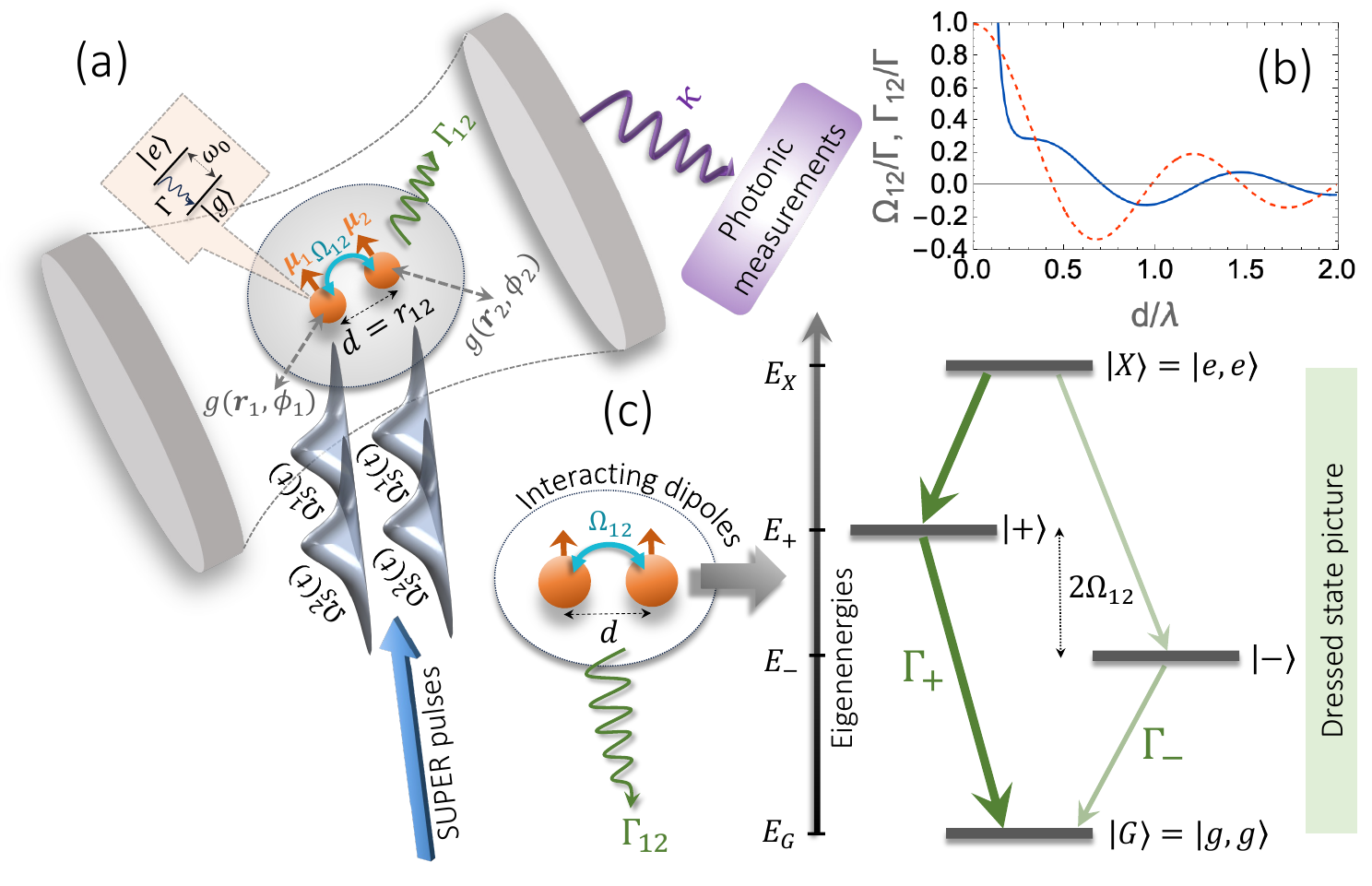}
\caption{Schematics for populating collective eigenstates of two interacting dipoles inside an optical cavity with SUPER pulses. (a) Two subwavelength-spaced QE (separation $|\bm{r}_{12}|$) with dipole moments $\bm{\mu}_i$ (perpendicular orientations to the separation vector are indicated by orange arrows) are placed inside a single-mode optical cavity of mode frequency $\omega_c$ and cavity decay $\kappa$. The system is transversely driven by sets of SUPER pulses ($\Omega^{1(2)}_{\text{S}}(t)$). The inset shows that each QE is a two-level system (ground state $|g\rangle$ and excited state $|e\rangle$) with transition frequency $\omega_0$ (wavelength $\lambda = 2\pi c/\omega_0$, where $c$ is the speed of light) and spontaneous emission rate $\Gamma$. The dipole-dipole interaction is mediated by the coherent coupling $\Omega_{12}=\Omega_{21}$ and the decay rates are $\Gamma_{ij}$ with $i,j\in\{1,2\}$ (where $\Gamma_{11} = \Gamma_{22} = \Gamma$ and $\Gamma_{12} = \Gamma_{21})$. The atom-cavity coupling for the $i^{\text{th}}$ QE is $g(\textbf{r}_i, \phi_i)$ with $\bm{r}_i$ the position vector and $\phi_i$ the relative phase of interaction. The cavity leakage $\kappa$ allows for photonic measurements in reality. (b) The variation  of collective energy shift $\Omega_{12}/\Gamma$ (blue-solid line) and collective decay rates $\Gamma_{12}/\Gamma$ (red-dashed line) for two parallel optical dipoles with inter-dipole separation $d/\lambda = |\bm{r}_{12}|/\lambda$. (c) The dressed state picture of the deep-subwavelength ($d \ll \lambda$) arrangement of two QEs results in eigenstates $\{|G\rangle,|\pm\rangle,|X\rangle\}$ (dressed states) with eigenenergies $E_G, E_{\pm}$ and $E_X$, respectively. Modified dissipative decay rates $\Gamma_{\pm} = \Gamma \pm \Gamma_{12}$ for the cascade transition channels $\ket{X}\Rightarrow|\pm\rangle\Rightarrow\ket{G}$ are displayed.}
\label{schematic}
\end{figure}
We consider two identical two-level quantum emitters (QEs) placed at subwavelength separation inside an optical cavity, which is transversely driven via SUPER pulses (see Fig.~\ref{schematic}(a)). The transition frequency and the transition dipole moment of each QE are $\omega_0$ and $\mu_0$, respectively. The spontaneous emission rate is given by $\Gamma$. The dipole-dipole interaction Hamiltonian can then be expressed as ($\hbar \equiv 1$)
\begin{align}
    \hat{H}_{\text{DD}} = \sum_{i}\omega_0\hat{\sigma}_i^+\hat{\sigma}_i^- + \sum_{i\neq j}\Omega_{ij}\hat{\sigma}_i^+\hat{\sigma}_j^-~,
    \label{eq:Hdd}
\end{align}
where the atomic raising operator is defined as $\hat{\sigma}_i^{+} = \ket{e}_{i}\!\bra{g}$ and the lowering operator as $\hat{\sigma}_i^{-} = (\hat{\sigma}_i^{+})^{\dagger}$. The collective coupling $\Omega_{ij}$ is a quantity depending on the inter-emitter separation $d = r_{ij} = |\bm{r}_{ij}| = |\textbf{r}_{i} - \textbf{r}_j|$, its spatial orientation $\hat{\bm{\varepsilon}}_{\bm{r}_{ij}} = \bm{r}_{ij}/r_{ij}$ and dipole orientation $\hat{\bm{\varepsilon}}_{\bm{\mu}_{i(j)}} = \bm{\mu}_{i(j)}/\mu_0$, where $\bm{r}_{i(j)}$ is the spatial position vector and $\bm{\mu}_{i(j)}$ the corresponding dipole vector of the $i(j)^{\text{th}}$ QE with $i,j\in\{1,2\}$ and $i\neq j$ (see Fig.~\ref{schematic}(a)). For two-level systems, the collective energy shift $\Omega_{ij}$ is given by~\cite{ficek:pr:2003}
\begin{align}
\Omega_{ij} =& -\frac{3\Gamma}{4} \bigg((1-\cos^2\theta) \frac{\cos{(k_0 r_{ij})}}{k_0 r_{ij}} \nonumber\\
-& (1 - 3\cos^2\theta) \bigg(\frac{\sin{(k_0 r_{ij})}}{(k_0 r_{ij})^2} + \frac{\cos{(k_0 r_{ij})}}{(k_0 r_{ij})^3}\bigg)\bigg)~,
\label{eq:coh}
\end{align}
where $k_0 = \omega_0/c$ is the wavenumber. In this article, we consider that the dipole moment vector of all QEs is aligned perpendicular to the $z$-axis, i.e. $\hat{\bm{\varepsilon}}_{\bm{\mu}_{i}} = \hat{\bm{\varepsilon}}_{i,\perp_z}$ along which the QEs are positioned and parallel (i.e., H-aggregate configuration~\cite{spano:cr:2018}). This arrangement leads to $\textbf{r}_{12} = d\cdot\hat{\bm{\varepsilon}}_z = -\textbf{r}_{21}$ and thus $\cos{(\theta)} = \hat{\bm{\varepsilon}}_{i,\perp_z}\cdot\hat{\bm{\varepsilon}}_z = 0$ which simplifies Eq.~(\ref{eq:coh}) to the following form
\begin{align}
    \Omega_{12} = -&\frac{3\Gamma}{4} \bigg( \frac{\cos{(k_0 d)}}{k_0 d} 
-  \bigg(\frac{\sin{(k_0 d)}}{(k_0d)^2} + \frac{\cos{(k_0 d)}}{(k_0d)^3}\bigg)\bigg)~,
\label{omega12}
\end{align}
with $\Omega_{12} = \Omega_{21}$ (see variation in Fig.~\ref{schematic}(b)). The $\hat{H}_{\text{DD}}$ (Eq.(\ref{eq:Hdd})) can be transformed into the dressed-state basis $\{\ket{G}=\ket{g,g},\ket{\pm},\ket{X} = \ket{e,e}\}$ with $\ket{\pm} = 1/\sqrt{2}(\ket{g,e} \pm \ket{e,g})$ and eigenenergies $E_G = 0$, $E_{\pm} = \omega_0 \pm \Omega_{12}$ and $E_X = 2\omega_0$ (see Fig.~\ref{schematic}(c) and further details are in Appendix-\ref{apen-dsp}). In the first excitation manifold, $\Omega_{12}$ lifts the degeneracy of the states $\ket{\pm}$. In this manuscript, we will focus our discussions on the symmetric ($|+\rangle$) and the antisymmetric ($|-\rangle$) branch, which illustrates the distance-dependent super and subradiant channels of the interacting two two-level QEs (see Appendix-\ref{apen-dsp} for details).

\subsection{SUPER Excitation Scheme}
\label{susec:SUPER}
The SUPER excitation scheme, first proposed theoretically~\cite{doris:PRXQ:2021} in 2021 and then experimentally demonstrated~\cite{karli:nl:2022,Super:arxiv:2023, boos:at:2024, joss:nl:2024}, is a rather unconventional scheme for exciting a QE from an energetically lower to a higher state, utilizing a carefully considered combination of two off-resonant Rabi oscillations with frequency modulation or using two far-detuned Gaussian pulses of $\sim$ ps time-scale. The Gaussian pulse SUPER scheme is referred to as the two-color (2C)-SUPER scheme~\cite{doris:PRXQ:2021}. The pulses overlap in time and result in a beat-like interference. Choosing ideal pulse parameter sets, one can realize a full population inversion to the excited state~\cite{doris:PRXQ:2021, karli:nl:2022, doris:PRB:2023}. The SUPER pulse can be described as follows
\begin{align}´
    \Omega_{\text{S}}(t) = \Omega_{\text{S}}^1(t)e^{-i\omega_1 t} + \Omega_{\text{S}}^2(t-\tau)e^{-i\omega_2t+i\varphi_X},
    \label{eq:SUPER_pulse}
\end{align}
where $\Omega^i_{\text{S}}(t) = (\alpha_i/\sqrt{2\pi\sigma_i^2}) \exp\{-t^2/(2\sigma_i)^2\}$ with $i$$\in$$\{1,2\}$ are the Gaussian pulses. The pulses are characterized by their pulse area $\alpha_i$, the temporal FWHM $\sigma_i$, detuning $\Delta_i = (\omega_i - \omega_0)$, a temporal shift between the pulses $\tau$ and a relative phase $\varphi_X$, leading to the SUPER parameter set: ($\Delta_1,\Delta_2,\sigma_1,\sigma_2,\alpha_1,\alpha_2,\tau,\varphi_X$). In Ref.~\cite{doris:PRXQ:2021}, it was shown that $\varphi_X$ does not alter the occupation of the final state, therefore, we choose $\varphi_X = 0$ for further discussion, which implies that the parameter set ($\Delta_1,\Delta_2,\sigma_1,\sigma_2,\alpha_1,\alpha_2,\tau$) will be optimized in later sections to ensure a close-to-unity population inversion in the excited states (collective and non-collective), i.e., $\ket{G} \Rightarrow \{\ket{\pm},\ket{X}\}$.

\subsection{Two Interacting Two Level QEs Inside an Optical Cavity}
\label{susec:Cavity}
Initially, the system is considered to be in the ground state $\ket{g,g}$. Now the deep-subwavelength QEs experience dipole-dipole interaction $\hat{H}_{\text{DD}}$ (Eq.~(\ref{eq:Hdd})).
We illuminate the system with two identical sets of SUPER pulses consisting of two time-overlapping Gaussian pulses, where each set individually targets one QE respectively (see Fig.~\ref{schematic}(a)). Exciting each QE with a separate SUPER pulse will allow for a flexible addressing of system states. Further discussions are in  Appendix-\ref{apen-dsp}, Sec.~\ref{susec:hybrid-states} and Appendix-\ref{apen-phase}. We consider the system to be inside a single-mode optical cavity to the system (Fig.~\ref{schematic}(a)), and we can write the following time-dependent Hamiltonian
\begin{align}
    \hat{H}_{\text{tot}}(t) = \hat{H}_{\text{DD}} + \hat{H}_{\text{C}} \underbrace{-\frac{1}{2}\sum_i(\Omega_{\text{S}}(t)\hat{\sigma}_i^+e^{i\vartheta_i} + \text{h.c.})}_{\hat{H}_{\text{SUPER}}}~.
    \label{Htot}
\end{align}
Without loss of generality we set $\vartheta_2 = 0$ and for simplicity we define $\vartheta_1 = \vartheta$ with $\vartheta\in(-\pi,\pi]$, which is the relative optical phase~\cite{lodhal:science:2023,van2025resonant}. This allows a straightforward discussion on addressing the two branches (symmetric and antisymmetric) in Fig.~\ref{schematic}(c) arbitrarily (see Sec.~\ref{susec:hybrid-states} for details). For example, $\vartheta = 0$ selectively couples to the $\ket{+}$ branch while $\vartheta = \pi$ targets the $\ket{-}$ branch. The Hamiltonian for a single-mode optical cavity and the emitter-field coupling is given by
\begin{align}
   \hat{H}_{\text{C}} = \omega_c\hat{a}^{\dagger}\hat{a} + \sum^2_{i=1}(g(\textbf{r}_i,\phi_i)\hat{\sigma}_i^+\hat{a} + \text{h.c.})~,\label{eq:HC}
\end{align}
where $\omega_c$ denotes the cavity frequency, $\hat{a}^{\dagger}$ ($\hat{a}$) are the creation (annihilation) operators of the cavity mode and $g(\textbf{r}_i,\phi_i) = g(\textbf{r}_i)\exp(i\phi_i)$ are the distance- and phase-dependent ($\phi_i\in(-\pi,\pi]$) cavity couplings. In a deep-subwavelength arrangement both QEs experience the same coupling strength, i.e., $|g(\textbf{r}_1,\phi_1)|$ = $|g(\textbf{r}_2,\phi_2)|$ = $g\in\mathbb{R}$. Analogous to $\vartheta_i$, the phase factors $\phi_i$ determine how the cavity mode couples to the system. The phases $\phi_i$ are induced via back-scattering processes in the cavity~\cite{painter:pra:2007}, for instance, due to surface roughness, which results in the coupling of different branches. The relative phase $\vartheta$ (Eq.~(\ref{Htot})), on the other hand, is an optically (externally) tunable quantity~\cite{lodhal:science:2023, van2025resonant}. As a result, the cavity mode couples purely to the symmetric branch $\ket{+}$ when $\phi_1 = 0 = \phi_2$ and equally couples to both the $\ket{+}$ and $\ket{-}$ branches when $\phi_1 = 0$ and $\phi_2 = \pi/2$. 
To reduce computational complexity, Eq.~(\ref{Htot}) is transformed into the rotating frame of $\omega_1$, i.e.
\begin{align}
        \hat{H}(t) = &-\Delta_1\sum_{i}\hat{\sigma}_i^+\hat{\sigma}_i^- + \sum_{i\neq j}\Omega_{ij}\hat{\sigma}_i^+\hat{\sigma}_j^- \nonumber\\
        &-\Delta_c\hat{a}^{\dagger}\hat{a} + \sum_i(g(\textbf{r}_i,\phi_i)\hat{\sigma}_i^+\hat{a} + \text{h.c.})\nonumber\\
        &- \frac{\Omega_{\text{S}}^1(t)}{2}\sum_i(\hat{\sigma}_i^+e^{i\vartheta_i} + \text{h.c.})\nonumber\\
        &- \frac{\Omega_{\text{S}}^2(t-\tau)}{2}\sum_i(\hat{\sigma}_i^+e^{i\vartheta_i + i(\Delta_1 - \Delta_2)t} + \text{h.c.})
        \label{eq:Ham1}~,
\end{align}
with $\Delta_c = \omega_c - \omega_1$. Under the Born-Markov approximation, the system's dynamics is governed by the master equation
\begin{align}
    \partial_t\hat{\rho}(t) = -i[\hat{H}(t),\hat{\rho}(t)] + \hat{\mathcal{L}}_{\text{Coll}}[\hat{\rho}(t)] + \hat{\mathcal{L}}_{\text{C}}[\hat{\rho}(t)]~,
    \label{master}
\end{align}
where $\hat{\rho}$ is the density matrix. The collective dissipation of the open quantum system is subject to the following Liouvillian super-operator
\begin{align}
    \hat{\mathcal{L}}_{\text{Coll}}[\hat{\rho}] = \sum_{ij}\frac{\Gamma_{ij}}{2}(2\hat{\sigma}_i^{-}\hat{\rho}\hat{\sigma}_j^{+} - \{\hat{\sigma}_j^{+}\hat{\sigma}_i^-,\hat{\rho}\})~,
\end{align}
where $\{\cdot,\cdot\}$ indicates the anti-commutator and $\Gamma_{ij}$ are the dipole orientation and distance-dependent collective decay rates. The collective decay rates are given by~\cite{ficek:pr:2003}
\begin{align}
\Gamma_{ij} = &\frac{3\Gamma}{2} \bigg((1-\cos^2{\theta}) \frac{\sin{(k_0 r_{ij})}}{k_0 r_{ij}} \nonumber\\
+& (1 - 3\cos^2{\theta}) \bigg(\frac{\cos{(k_0 r_{ij})}}{(k_0 r_{ij})^2} - \frac{\sin{(k_0 r_{ij})}}{(k_0 r_{ij})^3}\bigg)\bigg)~,
\end{align}
which, for the same dipole orientations as in Eq.~(\ref{eq:coh}), simplifies to
\begin{align}
\Gamma_{12} = &\frac{3\Gamma}{2} \bigg(\frac{\sin{(k_0d)}}{k_0 d}+ \bigg(\frac{\cos{(k_0d)}}{(k_0d)^2} - \frac{\sin{(k_0 d)}}{(k_0 d)^3}\bigg)\bigg)~,
\end{align}
with $\Gamma_{12} = \Gamma_{21}$ (see Fig.~\ref{schematic}(b) for the variation with $d/\lambda$). Similarly, the cavity loss is accounted for by the following Liouvillian
\begin{align}
    \hat{\mathcal{L}}_{\text{C}}[\hat{\rho}] = \frac{\kappa}{2}\bigg(2\hat{a}\hat{\rho}\hat{a}^{\dagger} - \{\hat{a}^{\dagger}\hat{a},\hat{\rho}\}\bigg)~.
\end{align}
This theoretical description of the coupled dipole system, embedded within a cavity and excited by the SUPER scheme, provides an opportunity for fundamental theoretical investigation, as detailed below.

\section{Selective Preparation of the Collective Eigenstates Using the SUPER Excitation Scheme}
\label{sec:POP}
The SUPER scheme offers many different advantages that conventional excitation schemes fail to provide \cite{doris:PRXQ:2021, karli:nl:2022, Super:arxiv:2023, doris:PRB:2023, boos:at:2024, Bracht:OQ:23, joss:nl:2024, Schumacher:PRR:2024}. Firstly, it is an off-resonant excitation scheme, and therefore, spectral separation of the pumping laser and the resulting emission is, in principle, trivial. This is especially true for spectroscopy of collective states. One of the advantages are non-unique parameter sets (see Sec.~\ref{susec:SUPER}). For simplicity, since the SUPER parameter set ($\underline{\Delta_1},\underline{\Delta_2},\underline{\sigma_1},\underline{\sigma_2},\alpha_1,\alpha_2,\tau$) consists of seven adjustable parameters, we choose to fix the underlined parameters. The parameters are provided in SI units and also in dimensionless units (with respect to $\Gamma,\lambda$, where $1/\Gamma = 1$ ns) as detailed below in Table-\ref{tab1}.
\begin{table}[h!]
\caption{\label{tab1}
The fixed pulse parameter sets for the $i^{\text{th}}$ pulse, where $i = 1, 2$: detuning ($\Delta_{1(2)}$) and FWHM ($\sigma_i$) in SI and scaled units (s.u.), respectively.}
\begin{ruledtabular}
\begin{tabular}{ccccc}
Units & $\Delta_1$ & $\Delta_2$ &  $\sigma_1$ & $\sigma_2$\\
\hline
SI & $-5$ meV & $-10$ meV & $6$ ps & $6$ ps \\
s.u. & $-7595.58$ & $-15191.16$ & $0.006$ & $0.006$
\end{tabular}
\end{ruledtabular}
\end{table}
The pre-adjusted parameters reduce the full variable space $\mathbb{R}^{7}$ to $\mathbb{R}^{3}$. Theoretically, one can choose the fixed parameters arbitrarily, however, the respective quantities should be easily adjustable experimentally. If not specifically mentioned, we set $\Gamma\tau = 0$, thus, we obtain the variable space $\mathbb{R}^2$. In Sec.~\ref{susec:POP}, we set $\Gamma\tau = 0.004$ for achieving near-unity inversion in state $\ket{-}$. Finding two suitable choices for $\alpha_1$ and $\alpha_2$ to populate a certain state $\ket{i}\in\{\ket{\pm}, \ket{X} = \ket{e,e}\}$ can be done rather easily by computing the final population as follows
\begin{equation}
    P_i^{\text{end}} = \text{Tr}(\ket{i}\bra{i}\hat{\rho}(t_{\text{end}}))~,
\end{equation}
where $t_{\text{end}}$ denotes the end of pulse duration in the SUPER mechanism.

In the following section, we first investigate the influence of a cavity environment (in Sec.~\ref{susec:POP}) and discuss possible effects of environmental decoherence (in Sec.~\ref{susec:DEC}) on the preparation of collective states. Subsequently, we disregard the optical cavity interface and discuss the optimized SUPER parameter sets that lead to close-to-unity population inversions in collective states in Sec.~\ref{inversion-nocavity}. Lastly, we study the preparation of hybrid collective states relying on the tunability of relative optical phases ($\vartheta$) in Sec.~\ref{susec:hybrid-states}.

\begin{figure*}[t]
\centering
\includegraphics[width=\linewidth]{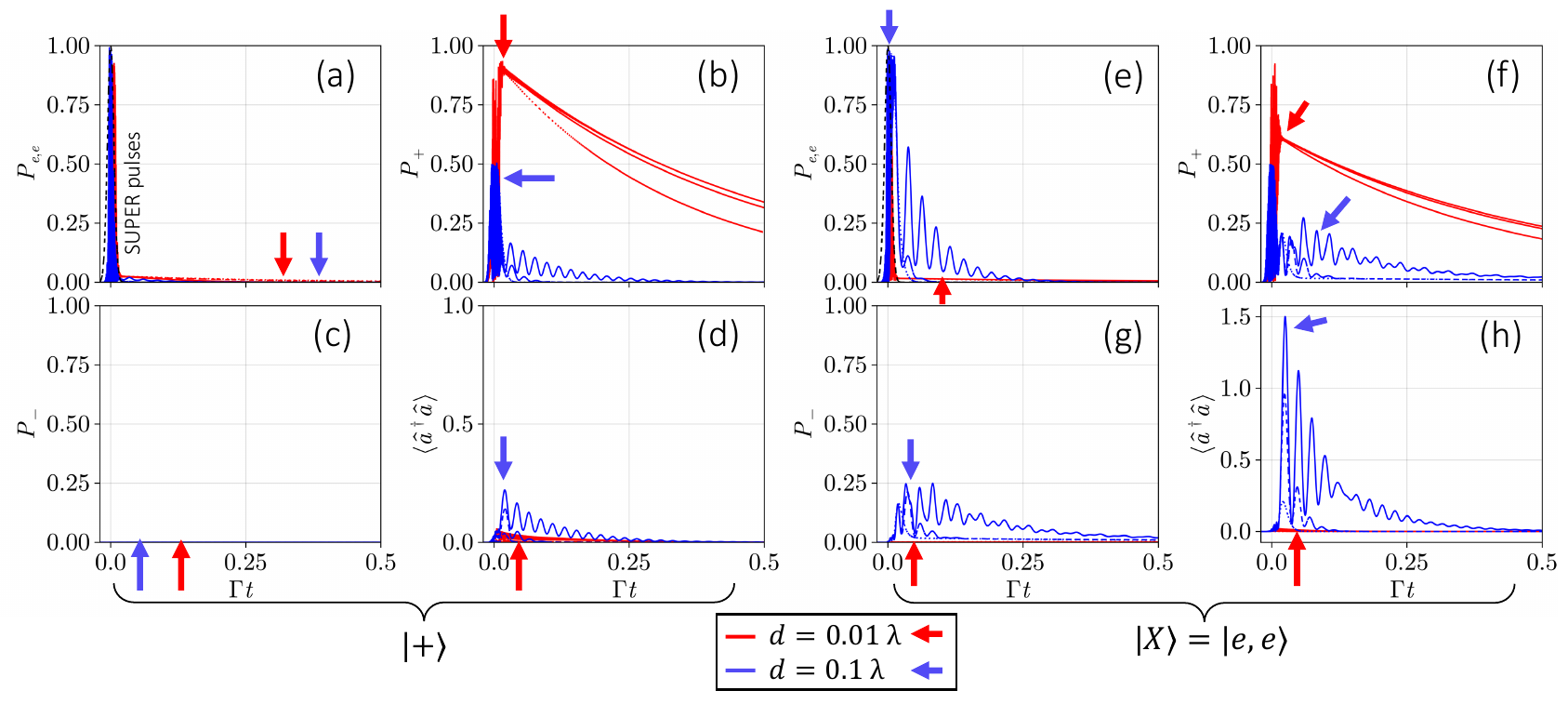}
\caption{Population dynamics for interacting subwavelength-spaced two QEs inside an optical cavity. The superradiant state $|+\rangle$ is populated with the SUPER pulses using the pulse areas $\alpha_1=68.25\pi$ and $\alpha_2 = 59.05\pi$ (plots (a)-(d), optimized for the highest inversion when $d/\lambda = 0.01$). We address $\ket{X} = \ket{e,e}$ with $\alpha_1=51.74\pi$ and $\alpha_2 = 70.48\pi$ (plots~(e)-(h), optimized for the highest inversion $d/\lambda = 0.1$). The blue curves represent $d = 0.1\lambda$ and the red curves refer to $d = 0.01\lambda$, respectively. The cavity parameters $g = 100~\Gamma$ and $\kappa\in g\cdot\{1/5,1,5\}$ are  represented by the line types $\{\text{solid},\text{dashed},\text{dotted}\}$ for each $d$ value. (a) and (e) show the population of $\ket{X}$, i.e., $P_{e,e}$, together with the overlapping pulse profiles (black dashed/dotted lines). (b) and (f) display the population of the $\ket{+}$ state ($P_+$).  In (c) and (g), the $\ket{-}$ population ($P_-$) are presented. The intra-cavity photon numbers ($\langle \hat{a}^{\dagger} \hat{a} \rangle$) can be seen in (d) and (h).}
\label{fig2}
\end{figure*}

\subsection{Population Inversion into Collective States in an Optical Cavity}
\label{susec:POP}
We position our dipole-interacting QEs inside an optical cavity (Fig.~\ref{schematic}(a)) where the interaction with the cavity mode (Eq.~(\ref{eq:Ham1})) can be controlled by adjusting the cavity coupling strength $g$ and the cavity-coupling phases $\phi_i$ ($i\in \{1,2\}$). The dissipation of the system is governed by the Lindblad master equation (Eq.~\ref{master}), when the system is weakly coupled to the environment. In Fig.~\ref{fig2}, we demonstrate how the inter-dipole separation $d$, the SUPER phase $\vartheta$~\cite{lodhal:science:2023,van2025resonant}, and the cavity coupling phases $\phi_i$ affect the system's population dynamics. The cavity mode is resonant to the atomic transition frequency, i.e $\Delta_c = \Delta_1$. We consider the following two cases. In Fig.~\ref{fig2}(a)-(d), we investigate the case of efficiently populating $\ket{+}$ for $d = 0.01\lambda$ (red lines) and $\phi_1 = 0 = \phi_2$ (here the cavity couples to the $\ket{+}$ branch only). Fig.~\ref{fig2}(e)-(h) display the case of efficiently populating $\ket{e,e}$ state at $d = 0.1\lambda$ (blue lines) with $\phi_1 = 0$ and $\phi_2 = \pi/2$ (here the cavity couples to both $\ket{\pm}$ branches equally). We choose $\vartheta = 0$ in Fig.~\ref{fig2}, because, on the one hand, we know that for $\vartheta \neq 0$ the SUPER mechanism couples partially (for $\vartheta = \pi$ fully) to the $\ket{-}$ branch, resulting in less population inversion of $P_+$ in the first case. On the other hand, $\vartheta = 0$ allows for an estimation of excess non-zero $P_-$ population stemming from a close-to-resonant cavity coupling in the second case (Fig.~\ref{fig2}(e)).

First, we focus on state preparation. In Fig.~\ref{fig2}(a)-(c), we observe that $P_{e,e}$ (red/blue curve) remains small after the SUPER excitation, while $P_+ > 91\%$ (red curve) is achieved for $d = 0.01\lambda$. Increasing the separation to $d = 0.1\lambda$ (blue line) implies a strong decrease of $\Omega_{12}$ (Fig.~\ref{schematic}(b)) and reduces the energy shift of $\ket{+}$ , i.e., $E_+\approx -\Delta_1$, leading to a decrease of $P_+$ to $P_+ \ll 25\%$ with zero population in $\ket{-}$. Similarly, Fig.~\ref{fig2}(e)-(g) we obtain $P_{e,e}>95\%$ for $d = 0.1\lambda$ (blue line). The population decreases to $P_{e,e}\ll10\%$ for $d = 0.01\lambda$ (red line), owing to the fact that the excitation process hinges on a large energy shift $E_{+}$ when we use identical individual SUPER pulses for each QE (see Appendix-\ref{apen-dsp}). We find $P_{+}>60\%$ for $d = 0.01\lambda$ (red line). The SUPER scheme is designed to invert in between two isolated energy levels. Thus, parameters optimized for a pair of energy levels may work reasonably well for other sets of discrete energy levels too. However, total probability,  $P_{e,e}+P_{+}+P_{-}+P_{G} = 1$, must remain conserved. We find $P_- \approx 0$ (red line) and $P_- >0$ (blue line). This is explained by looking at the cavity coupling. In Fig.~\ref{fig2}(a)-(d), the cavity couples to the $\ket{+}$ branch only. We find an oscillating photon number $\langle \hat{a}^{\dagger} \hat{a}\rangle$ (Fig.~\ref{fig2}(d)) for $d = 0.1\lambda$ (blue line), as with $E_{+} \approx -\Delta_1$ the photons couple resonantly to the cavity. This is not the case for $d = 0.01\lambda$, where $E_+\gg-\Delta_1$. This is also the case in Fig.~\ref{fig2}(e)-(h), however, here, we observe $P_->0$ because the cavity couples to both branches equally. Ideally, we expect a maximal value of two photons in Fig.~\ref{fig2}(h) for $d = 0.1\lambda$ due to the decay cascade $\ket{(e,e),0_{\mp},0_{\pm}}\Rightarrow\ket{\pm,1_{\mp},0_{\pm}}\Rightarrow\ket{(g,g),1_{\mp},1_{\pm}}$. However, only $\max_{\Gamma t}{\langle\hat{a}^{\dagger}\hat{a}\rangle} = 1.5$ can be achieved, because one photon is emitted earlier, yielding a non-zero probability to re-excite the system: $\ket{\pm,1_{\mp},0_{\pm}}\Rightarrow\ket{(e,e),0_{\mp},0_{\pm}}$.  Although here we choose $\Delta_c = \Delta_1$, in principle, one could also choose other cavity frequencies; those could result in different temporal evolution of the population of the prepared state, however, the respective maximum achievable population and robustness over different cavity coupling regimes would remain the same.

The swing-up mechanism, considering a single QE~\cite{doris:PRXQ:2021,karli:nl:2022}, implies an off-resonant coupling to the cavity during the excitation process~\cite{Schumacher:PRR:2024}. This stems from AC-Stark shifts~\cite{axt:prr:2020} induced by the SUPER pulses. Hence in Sec.~\ref{inversion-nocavity}, we will neglect the cavity when finding optimal parameters for an efficient inversion of the collective states. In the latter part of this article, cavity interface will be utilized for some photonic measurements.

Note that, the phases $\phi_i$ and $\vartheta$ possess an equivalent influence on accessing the bright (superradiant) or dark (subradiant) channel, as we are in the deep-subwavelength regime, both for the cavity and the SUPER coupling. In this section, as we are discussing the effect of the optical cavity, we vary $\phi_1, \phi_2$. In Sec.~\ref{susec:hybrid-states}, later, we will explore the possibility of accessing both dark and bright channels by changing only the relative phase $\vartheta$ and neglecting the cavity interface.

\begin{figure*}
\centering
\includegraphics[width=\linewidth]{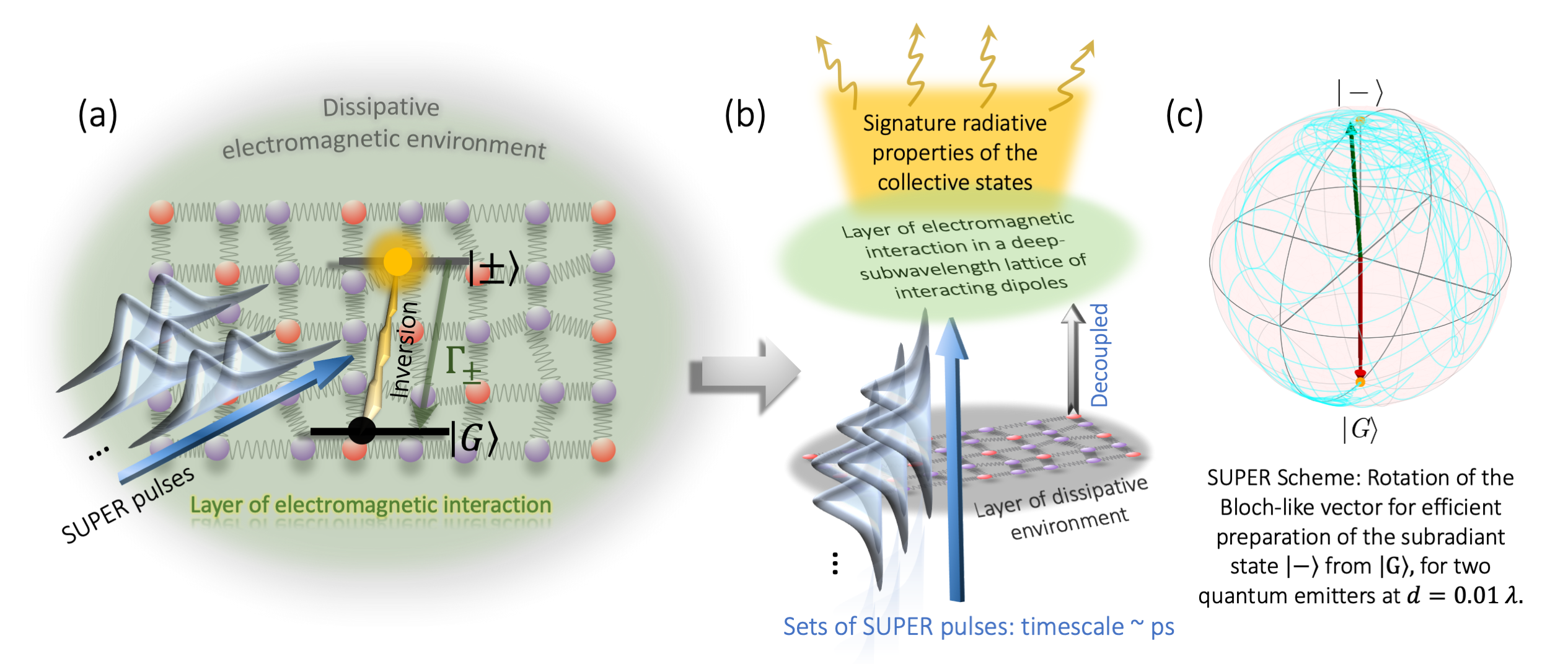}
\caption{(a) A schematic illustration of the efficient inversion by ultrashort SUPER pulses ($\sim$ ps) which drive the system from the ground state ($\ket{G}$) (black-circle) to one of the collective states ($\ket{\pm}$) (yellow-circle) of the QEs, with collective decay rates $\Gamma_{\pm}$. The interacting dipoles form a `layer of electromagnetic interaction' immersed in a dissipative environment, e.g., due to phonons (shown above as lattice vibrations in the background), having a timescale larger than the SUPER process or on-site energy imperfections (symbolically represented by different colors in the background lattice). A near-unity inversion can be accomplished despite the presence of the dissipative environment. (b) SUPER pulses can then selectively and separately (to a reasonable extent) access the pure layer of dipole-dipole interaction, which was earlier submerged into the dissipative environment and open opportunities to directly witness the signature collective radiative properties. (c) Dynamics of the density matrix corresponding to the sub-Hilbert space spanned by $\{\ket{G},\ket{-}\}$ (see. Sec.~\ref{susec:hybrid-states}), represented as Bloch-like vector (cyan-curve) for the $\ket{G}\Rightarrow\ket{-}$ transition, using the SUPER scheme in the absence of a cavity (pulse parameters are mentioned in the text). The red downward vector denotes the initial state $\ket{G}$, and the green upward vector represents the final $\ket{-}$ state.}
\label{schematic2}
\end{figure*}

\subsection{SUPER Excitation of Coupled Quantum Emitters Amidst Decoherence}
\label{susec:DEC}
Let us consider two QEs at deep-subwavelength separation experiencing strong dipole-dipole interaction in a noisy environment. The system is then being illuminated by sets of SUPER pulses to have a near-unity inversion in one of the collective states $\ket{+}$ or $\ket{-}$ (Fig.~\ref{schematic}(a)). Note that, the layer of pure electromagnetic interaction, which includes dipole-dipole interaction and corresponding collective decays, is in reality \emph{immersed} in a dissipative environment, which is an `ocean' of many frequencies and experiencing environmental triggers. Hence, the layer experiences finite environmental exchanges (see Fig.~\ref{schematic2}(a)). These exchanges include interaction with phonons (lattice vibrations) due to elevated temperature or from the on-site static energy imperfections of the QEs, as well as the influence of dephasing mechanisms via the substrate.

\subsubsection{Decoherence in an Optical Cavity}
\label{diss-cavity}
We already know that the SUPER scheme~\cite{doris:PRXQ:2021, karli:nl:2022, doris:PRB:2023} is a mechanism only between two discrete energy levels (for e.g., $\ket{G}\Rightarrow\ket{\pm}$ and $\ket{\pm}\Rightarrow\ket{X}$). Here we consider SUPER pulses of a duration of $\sigma_i$ = 6 ps (Table-\ref{tab1}). In Ref.~\cite{Bracht:OQ:23}, it has been shown that the excitation to an energetically higher state for a single quantum dot is only weakly influenced by phonons. This is a remarkable advantage over other coherent excitation schemes, for instance, two-photon absorption, where the achieved final population in the excited state decreases rapidly for higher temperatures~\cite{Bracht:OQ:23}. Intuitively, the SUPER scheme works fast, relies on off-resonant pulses, and importantly, traces an unconventional trajectory during the rotation of the Bloch vector on the Bloch sphere~\cite{doris:PRXQ:2021} and thereby induces an effective decoupling of the excitation (target state) from the environment~\cite{Bracht:OQ:23, Schumacher:PRR:2024}. In the coupled emitter case as discussed in Sec.~\ref{susec:POP}, we found a similar suppression of the environmental coupling when populating bright $\ket{+}$ states (at $d = 0.01\lambda$) with SUPER pulses (Fig.~\ref{fig2}(b)), which leads to very low population in the cavity mode (red curve in Fig.~\ref{fig2}(d)) in the presence of cavity-mediated system-environment interaction. The excited collective states $\ket{\pm}$ and also $\ket{X}$ experience an AC-Stark shift~\cite{axt:prr:2020}, and the respective Bloch-like vector (see. Sec.~\ref{susec:hybrid-states}) also traces an unconventional time-dependent trajectory on the Bloch sphere. As a result, they become off-resonant to the cavity mode, which drastically minimizes the possibility of re-excitation even in the strong system-cavity coupling regime. In short, the environmental exchanges of the pure `layer of electromagnetic interaction' are much reduced in an optical cavity.

\subsubsection{Decoherence from Static Position Disorders}
\label{diss-position}
Now, we map the above picture intuitively in the domain where QEs are experiencing position fluctuations in an environment/substrate. In particular, we disregard the cavity and consider the influence of lattice vibration (Fig.~\ref{schematic2}(a)) as on-site static position disorders (roughly). Here we consider a small fluctuation in the position of each QEs and the modified Hamiltonian in Eq.~(\ref{Htot}) can then be rewritten as
\begin{align}
    \hat{\tilde{H}}^{ph}_{\text{tot}}(t) =& \underbrace{\hat{\tilde{H}}^{ph}_{\text{DD}}(\epsilon_r,\tilde{r}_{ij})}_{\substack{\text{modified~} {\hat{{H}}_{\text{DD}}}\\ \text{for on-site}\\ \text{static position disorder}}}
    - \frac{1}{2}\sum_i(\Omega_{\text{S}}(t)\hat{\sigma}_i^+e^{i\vartheta_i} + \text{h.c.})~.
    \label{H-pos-dis}
\end{align}
The last term represents the SUPER pulses. For parallel dipoles (in H-configuration) (as in Fig.~\ref{schematic}(a)), we redefine Eq.~(\ref{eq:Hdd}) as follows
\begin{align}
    \hat{\tilde{H}}^{ph}_{\text{DD}} (\epsilon_r,\tilde{r}_{ij}) = \sum_{i}\omega_0\hat{\sigma}_i^+\hat{\sigma}_i^- + \sum_{i\neq j}\tilde{\Omega}_{ij} (\epsilon_r|\mathbf{\tilde{r}_i} - \mathbf{\tilde{r}_j}|)\hat{\sigma}_i^+\hat{\sigma}_j^-~.
    \label{HDD-dis}
\end{align}
Here, $\mathbf{\tilde{r}_i}$ is the mean position of the $i^{\mathrm{th}}$ QE (considering a normal distribution with width $\epsilon_r$ and averaged over many realizations) and $\tilde{r}_{ij} = |\mathbf{\tilde{r}_i} - \mathbf{\tilde{r}_j}|$ is the modified inter-emitter separation. In reality, $\epsilon_r$ could be roughly realized as a temperature-dependent quantity and a theoretically tunable parameter. We consider that the fluctuations around the position of the QE's are much smaller, i.e., $(|\mathbf{{r}_{i}}| - |\mathbf{\tilde{r}_{i}}|)/ |\mathbf{{r}_{i}}|\ll 1$. For a deep-subwavelength lattice, i.e. $d=0.01 \lambda$, the collective optical shifts are normally very large in magnitude, hence, one can write 
\begin{align}
    \frac{|\Omega_{ij} (|\mathbf{{r}_i} - \mathbf{{r}_j}|) - \tilde{\Omega}_{ij} (\epsilon_r|\mathbf{\tilde{r}_i} - \mathbf{\tilde{r}_j}|)|}{|\Omega_{ij} (|\mathbf{{r}_i} - \mathbf{{r}_j}|)|} \ll 1~.
\end{align}
This implies that at low temperature, i.e., with much slower lattice vibrations (on a timescale greater than ps), one can make the following reasonable approximation 
\begin{align}
    \hat{\tilde{H}}^{ph}_{\text{DD}} (\epsilon_r,\tilde{r}_{ij}) \simeq \hat{H}_{\text{DD}} (r_{ij})~.
\end{align}
Therefore, for slower lattice vibrations, the ultrashort SUPER scheme ($\sim$ ps) effectively accesses the pure collective states $\ket{\pm}$ of the interacting QEs (dynamics guided by Eq.~(\ref{master})) and then induces an AC-Stark shift to the energy levels and a rotation of the Bloch-like vector on the Bloch sphere, which effectively decouples the `layer of electromagnetic interaction' from the other modes of the dissipative environment which were near resonant before (following the similar principle as with the resonant cavity case, which we demonstrated earlier in Sec.~\ref{diss-cavity}) (see Fig.~\ref{schematic2}(b), (c)).  Subsequently, as the temperature rises, the excitation with the SUPER scheme will not be significantly impacted (in contrast to other coherent excitation methods), as was also demonstrated in Ref.~\cite{Bracht:OQ:23} with a single quantum dot-cavity system. However, to be really precise, one needs to take into account the whole spectrum of modes of the reservoir for a specific system. On a hopeful note, even if substantial modes are initially coupled to the pure `layer of electromagnetic interaction', the prepared collective state with SUPER pulses will always help the excitation to become decoupled from some of the modes of the reservoir, and thereby provide access to more robust signature radiation properties. 

Fig.~\ref{schematic2}(c) displays the trajectory of the Bloch-like vector (in the absence of a cavity and Sec.~\ref{susec:hybrid-states} later illustrates more on it) when addressing $\ket{G}\Rightarrow\ket{-}$ transition with SUPER pulses for two deep-subwavelength QEs. We utilize the parameters from Table-\ref{tab1}, $d = 0.01 \lambda$, $\vartheta = \pi$, $\Gamma\tau = 0.004$, pulse areas $\alpha_1=20\pi$ and $\alpha_2 = 40\pi$ to obtain around 99\% inversion in the $\ket{-}$ state.

\begin{figure*}
\centering
\includegraphics[width=\linewidth]{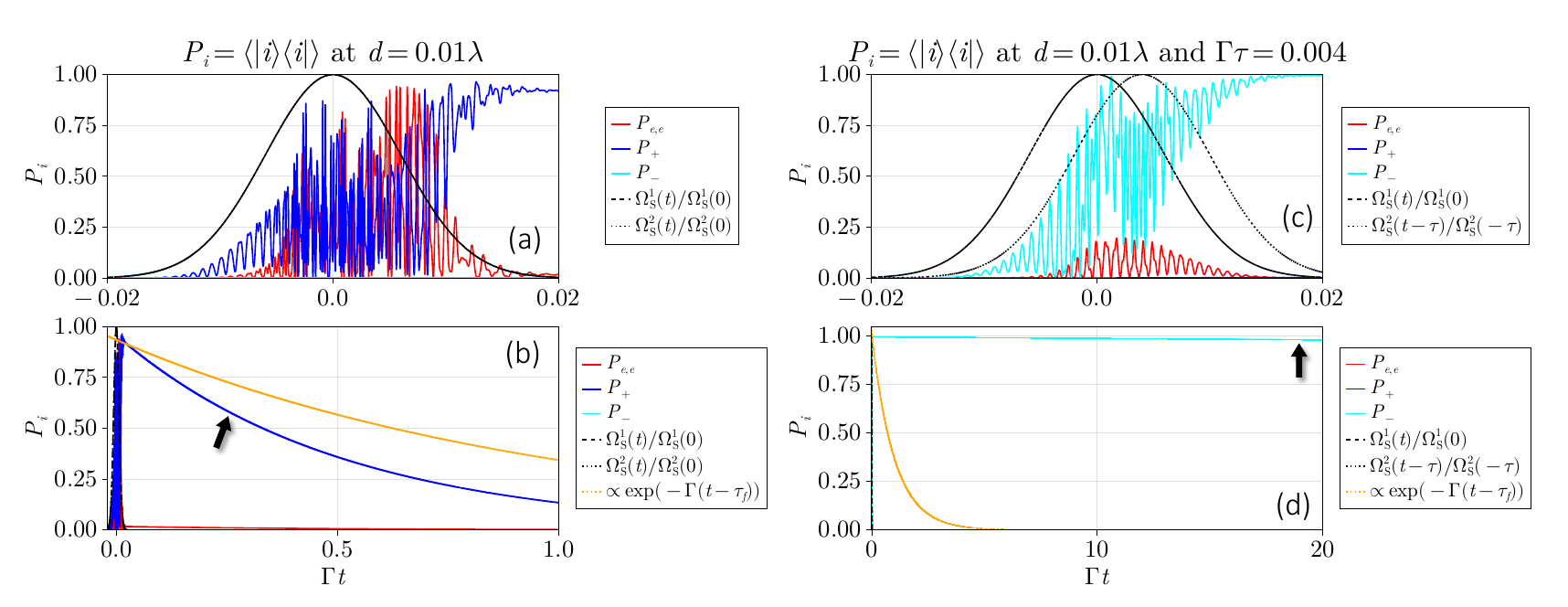}
\caption{Efficient preparation of collective eigenstates ($|\pm\rangle$) of deep-wavelength spaced ($d = 0.01\lambda$) optical dipoles using the SUPER scheme in the absence of an optical cavity. Gaussian pulse profiles $\Omega_{\mathrm{S}}^{1(2)}(t-\tau)/\Omega_{\mathrm{S}}^{1(2)}(0(-\tau))$ are represented with black-solid and black-dotted curves in plots (a),(b)) (here zero time-delay, $\Gamma \tau = 0$) and (c),(d) (here finite time-delay, $\Gamma \tau = 0.004$). Here, $\Omega_{\mathrm{S}}^1(2)(0(-\tau))$ simply denote the maximal values of the corresponding pulses. In particular, in plots (a) and (b), we address the superradiant $\ket{+}$ state with a SUPER phase $\vartheta = 0$ and pulse areas $\alpha_1=68.25\pi$ and $\alpha_2 = 59.05\pi$. A population inversion of $P_+>91\%$ in state $|+\rangle$ (blue-solid curve) is achieved. In plots (c) and (d), to address the subradiant $\ket{-}$ state, we use $\vartheta = \pi$ and pulse areas $\alpha_1=20\pi$ and $\alpha_2 = 40\pi$. This results in a population inversion of  $P_-> 99\%$ (cyan-solid curve) in state $\ket{-}$((c)-(d)). The decay of these prepared super- and subradiant state populations is shown on a longer timescale in plots (b) and (d) and indicated by black arrows. The orange line in plots (b) and (d) depicts the independent exponential decay with rate $\Gamma$ for comparison. The long-time evolution demonstrates a superradiant decay of $|+\rangle$ in (b) (blue-solid curve) and, importantly, a very long-lasting excitation storage characteristic, i.e.\ subradiance, for $\ket{-}$ (cyan-solid curve) in (d).}
\label{fig3}
\end{figure*}

\subsubsection{Decoherence due to On-site Energy Imperfection}
\label{diss-energy}
As discussed in the section above, equivalently, one could also consider the effect of on-site static energy disorder (i.e., non-identical energy levels) in the QEs (instead of cavity interaction (Sec.~\ref{diss-cavity}) or position disorder (Sec.~\ref{diss-position}) of the QEs). This is particularly common for solid-state emitters or molecules when embedded in a substrate. As discussed in Ref.~\cite{gauger:nc:2014}, the on-site energy disorders (inhomogeneous broadening) can be accounted for via a modified Hamiltonian~\cite{book:may:kuhn,Rebentrost:NJP:2009}
\begin{align}
    \hat{\tilde{H}}^e_{\text{tot}}(t) = \sum^2_{\nu=1} \omega^{\text{e}}_\nu \hat{\sigma}_{\nu}^+\hat{\sigma}_{\nu}^- &+ \sum_{i\neq j}\Omega_{ij} \hat{\sigma}_i^+\hat{\sigma}_j^-\nonumber\\
    -& \underbrace{\frac{1}{2}\sum_i(\Omega_{\text{S}}(t)\hat{\sigma}_i^+e^{i\vartheta_i} + \text{h.c.})}_{\hat{H}_{\text{SUPER}}}~,
    \label{H-en-dis}
\end{align}
where $\nu$ represents the site number and $\omega^{\text{e}}_{\nu}$ is the site energy, which varies due to the dissipative environment. This could be modeled as an average over many realizations of a normal distribution of width $\epsilon_e$ around atomic frequency $\omega_0$; i.e., $\omega^{\text{e}}_\nu = \epsilon_e \omega_0$. This will also provide an equivalent description of the system-environment interaction as we had utilized earlier with the cavity mode (in Sec.~\ref{susec:POP}).

In Ref.~\cite{gauger:nc:2014}, the authors theoretically demonstrated the robustness (up to a reasonable extent) of the effect of energy disorder on superradiance and superabsorption. For very high disorder, the pure bright and dark collective states would start mixing. Thus, for small (reasonable) changes in the site energy, i.e., roughly when
\begin{equation}
    \epsilon_e = \frac{\omega^{\text{e}}_{\nu}}{\omega_0} \lesssim \Gamma~,
\end{equation}
the Hamiltonian in Eq.~(\ref{H-en-dis}) will prohibit mixing of pure bright and dark states (see, for instance, Fig. F1 in Ref.~\cite{pal:njp:2025} in the context of bio-inspired nanoscale ring geometry). The above limit allows for pure collective states to retain their unique optical signatures. With reasonably less disorder, therefore for superradiance, the effective linewidth of the collective states will remain increased ($\Gamma+\Gamma_{12}$), i.e., exhibit faster decay and for subradiance it will remain narrowed down ($\Gamma-\Gamma_{12}$), i.e., showcase lesser decay (of course when compared to the case of independent QE). The prepared states also exhibit robustness against environmental triggers (up to an extent) for the SUPER excitation process (as happens in the presence of the cavity interface). To illustrate once again, following our previous insights, the efficiently prepared collective states will eventually experience an AC-Stark shift when prepared by SUPER pulses, and the Bloch-like vector traces an unconventional rotation on the Bloch sphere (similar to Fig.~\ref{schematic2}(c)), and as a result, become off-resonant to the environmental effects (reasonably). The excitation will therefore remain effectively decoupled from the dissipative environment caused by site energy imperfections (up to an extent), which allows for the system to showcase its signature radiative properties (as shown in Fig.~\ref{schematic2}(b)).

\begin{figure*}
\centering
\includegraphics[width=\linewidth]{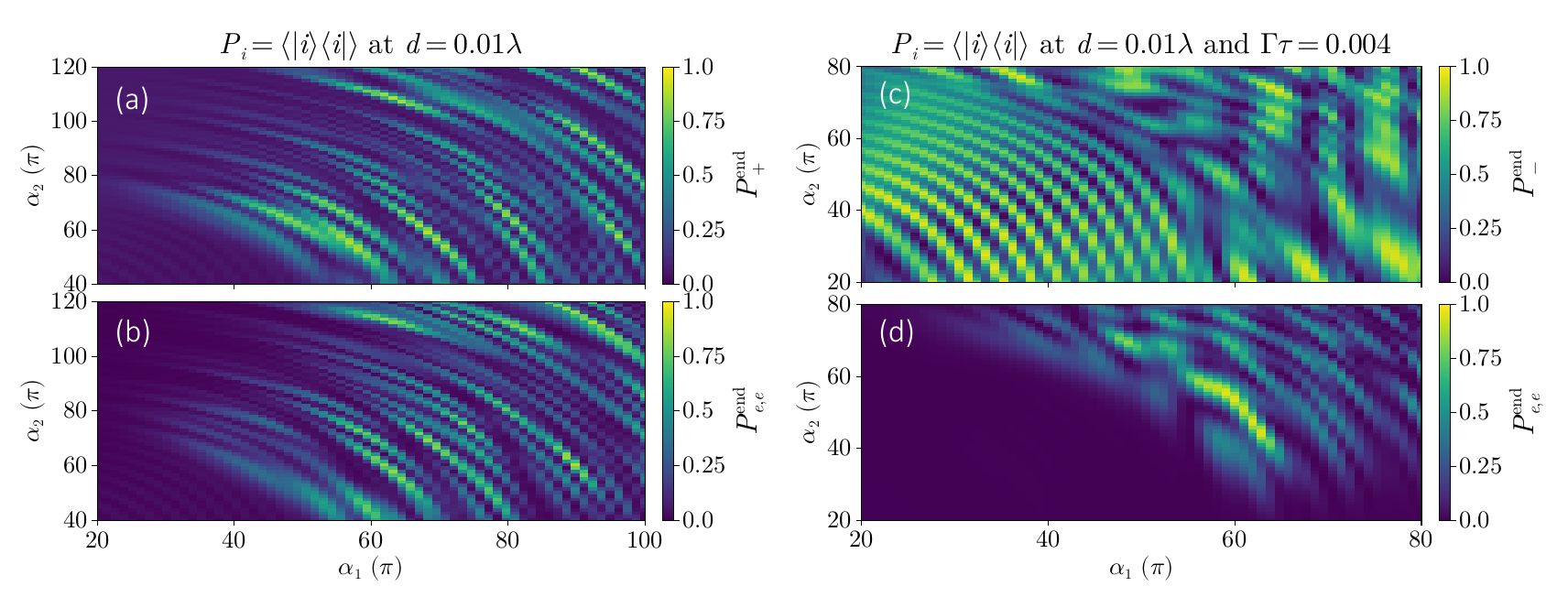}
\caption{Heatmaps visualizing pulse parameters $(\alpha_1,\alpha_2,\Gamma\tau)$ for highest possible inversions in state population with $\ket{i}\in\{\ket{+}, \ket{-}, \ket{e,e}\}$ and without the cavity. Parameters $(\Delta_1,\Delta_2,\sigma_1,\sigma_2)$ are fixed (see Table-\ref{tab1}). For (a)-(b), the final populations $P^{\mathrm{end}}_i = \langle\ket{i}\bra{i}\rangle$ are plotted for $\ket{i}\in\{\ket{+}, \ket{e,e}\}$ with $\alpha_1\in[20,100],\alpha_2\in[40,120]$ and $\Gamma\tau = 0$. Bright areas indicate high population inversion. For (c)-(d), $P^{\mathrm{end}}_i = \langle\ket{i}\bra{i}\rangle$ are depicted for $\ket{i}\in\{\ket{-},\ket{e,e}\}$ with $\alpha_1\in[20,80]$,$\alpha_2\in[20,80]$. The $\ket{-}$ branch is successfully addressed using a SUPER phase of $\vartheta = \pi$ and a finite time-shift of $\Gamma\tau = 0.004$. In the context of the preparation of collective states, plots (a) and (c) are of primary interest. Note, that we obtain a maximal final population $P_+ = \max_{\alpha_1,\alpha_2}P_+^{\text{end}} > 91\%$ for the parameter set $(\alpha_1 = 68.25\pi,\alpha_2=59.05\pi)$ and $P_- = \max_{\alpha_1,\alpha_2}P_-^{\text{end}}>99\%$ for the parameter set $(\alpha_1 = 20\pi,\alpha_2=40\pi)$, as shown in Fig.~\ref{fig3}.}
\label{heatmaps-collective}
\end{figure*}

\subsubsection{Further Possibilities}
Looking ahead, we expect the qualitative features discussed above might remain the same for very deep subwavelength-spaced lattices of quantum emitters, in particular, molecular arrangements in biological systems~\cite{nori:np:2013, Kohler:revbio:2006} to investigate more clearly the presence and/or influence of the collective states~\cite{vangrondelle:jpcb:1997, ferrari:jpcc:2014} in light harvesting mechanism (although, note that system-environment interaction is unfortunately too complex in this case to explain solely relying with Markovian master equation approach as detailed in Ref.~\cite{pal:njp:2025}) or synthetic molecular configurations~\cite{Anderson:nc:2022} (system-environment interaction is fortunately more controllable here~\cite{gauger:prl:2025}). In these cases, or more generally, using a single set of SUPER pulses (in particular, only two pulses) could be experimentally more feasible for preparing collective states, amidst a reasonable decoherence in the environment. It may also mitigate the effects of environmental dissipation up to a certain extent, allowing one to isolate (reasonably) and study the pure layer of electromagnetic interaction, governed by the dipole-dipole interaction. We intend to address this possibility in the future. In this context, it's worth noting that molecular aggregates are known to be resilient (upto a certain extent) to imperfections, such as inhomogeneous broadening~\cite{mukamel:jcp:1989} and dephasing~\cite{mukamel:pra:1988}. If this were true, this venture to efficiently and selectively prepare collective states with SUPER scheme would be crucial for understanding the importance of the biological geometry purely relying on quantum optical Hamiltonians for exploring the landscape of bio-inspired optically efficient configurations~\cite{gauger:nc:2014, gauger:2024:spie, gauger:prx:2023, cardoner:oe:2022, Holzinger:24, pal:njp:2025}. However, these insights will nevertheless require more detailed and in-depth future investigations for confirmation.

\subsection{Parameters for Cavityless Inversions in Bright and Dark Collective States}
\label{inversion-nocavity}
In this section, we discuss and detail the optimized SUPER parameter sets for obtaining a high population inversion in the states $\ket{i}\in\{\ket{\pm}, \ket{X}\}$. We remove the cavity from our model, as the SUPER excitation scheme is robust to the choice of cavity coupling strength (as discussed in Sec.~\ref{susec:Cavity}). Fig.~\ref{fig3} (a), (c) show the swing-up behavior without a cavity for efficient preparation of the collective states $\ket{\pm}$ of two-interacting two-level QEs at deep subwavelength separation (Fig.~\ref{schematic}(c)), i.e.\ $d = 0.01 \lambda$. With optimal pulse parameters, we can achieve approximately $91\%$ population for the $|+\rangle$ state (Fig.~\ref{fig3}(a), blue-solid curve) and over $99\%$ for the $|-\rangle$ state (Fig.~\ref{fig3}(c), cyan-solid curve). For both cases, the final population in $\ket{X} = \ket{e,e}$ is near zero. We display the long-term dynamics after the selective preparation of the collective states $\ket{\pm}$ as well (Fig.~\ref{fig3}(b), (d)). The bright state (blue-solid curve) decays faster (Fig.~\ref{fig3}(b)) and the dark state (Fig.~\ref{fig3}(d)) (cyan-solid curve) decays much slower than the independent decay of the individual QEs (orange-dotted curve). It is important to note that the SUPER mechanism, therefore, allows one to access bright and dark states selectively (this is crucial as dark states are, in general, decoupled from the electromagnetic environment) by a suitable choice of externally tunable pulse parameters only. This allows for a direct preparation of a desired collective state. Looking at the existing experimental studies dealing with solid-state emitters and molecules~\cite{Gerardot:sc:2022, lounis:nc:2022, lodhal:science:2023, Hood:np:2024}, our above results might motivate new approaches to selective preparation of collective states in interacting QE systems.

It is important to note that the SUPER parameter sets are not unique, as displayed in Fig.~\ref{heatmaps-collective} and also previously demonstrated in literature when dealing with single quantum dots~\cite{Schumacher:PRR:2024, doris:PRXQ:2021, Bracht:OQ:23}. This implies that the final probabilities $P_+^{\text{end}}, P_-^{\text{end}}$ are by no means rigorous upper bounds. In this section, we explicitly write $P_i^{\text{end}}$ instead of $P_i$, pointing out that we focus on the final population immediately after the SUPER procedure. For the present article, we consider $P_+^{\text{end}}>91\%$ and $P_-^{\text{end}}>99\%$ to be sufficient. In Fig.~\ref{heatmaps-collective} we display the heatmaps with optimized parameters of the final population $P_+^{\text{end}}, P_-^{\text{end}}$ and $P_{e,e}^{\text{end}}$ of $\ket{+},\ket{-}$ and $\ket{e,e} = \ket{X}$ at $d = 0.01\lambda$, respectively (we focus mainly on achieving an as high as possible population inversion for $\ket{\pm})$. We use a $64\times64$ parameter domain, that is, $64\times64$ combinations of $\alpha_1 \in[20,100]$ and $\alpha_2 \in[40,120]$ for $\ket{+}$ state and of $\alpha_1 \in[20,80]$ and $\alpha_2 \in[20,80]$ for $\ket{-}$ state, chosen equidistantly.

Furthermore, the SUPER parameters are highly sensitive to drastic changes of the collective energy shifts $\Omega_{12}$ (Eq.~\ref{omega12}) of the levels, which depend on the distance $d$. Hence, if one were to choose a different subwavelength separation, e.g., $d = 0.1\lambda$, then, $\Omega_{12}$ would  change and imply a significant decrease in the collective energy shift (see Fig.~\ref{schematic}(c)), leading to near degenerate energies $E_{\pm} = -\Delta_1 \pm \Omega_{12} \sim -\Delta_1$. As a result, one would need to suitably optimize the parameters. It turns out that both $\ket{+}$ and $\ket{-}$ states can no longer be populated efficiently when $d \gg 0.01\lambda$ and $\vartheta = 0$ or $\vartheta = \pi$, respectively. Due to the near-degenerate nature of the collective states and duration time of the very short SUPER pulses, populations around $P_{\pm}^{\text{end}}<50\%$ can only be achieved.

The above observations for efficient inversions into the respective superradiant and subradiant states via SUPER pulses are expected to remain qualitatively similar, even with changes in the dipole orientations and in any deep-subwavelength separation of the QEs, as long as the energies $E_{\pm}$ are approximately the same, i.e., $\Omega_{12}^{\text{mod}} \approx \Omega_{12}$, where $\Omega_{12}^{\text{mod}}$ denotes the energy shift with respect to the modified dipolar orientation and sub-wavelength separation (the states $\ket{\pm}$ may swap the nature of super- and subradiance, implying $E_{\mp}^{\text{mod}}\approx E_{\pm}$). Head-to-tail configuration, i.e., J-configuration~\cite{spano:cr:2018}, for instance, where radiative properties will change, may also be addressed with the same parameters (if the above statement is fulfilled).
\begin{figure*}
\centering
\includegraphics[width=\linewidth]{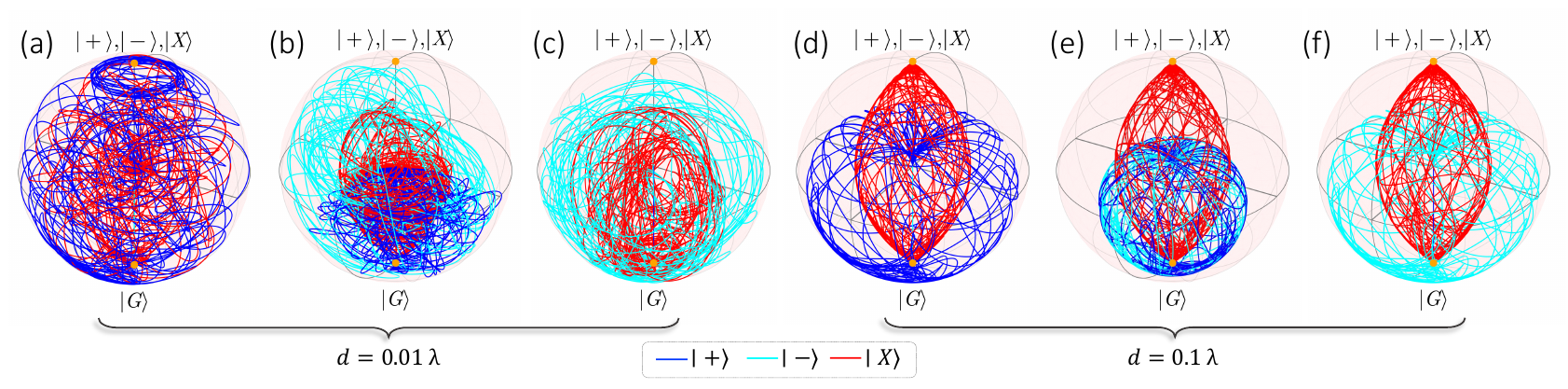}
\caption{In plots ((a)-(f)), blue trajectories on the Bloch spheres represent the population of the superradiant $\ket{+}$ state excited from $\ket{G}$ over time, i.e., $P_+$ with the relative phase $\vartheta\in\{0,\pi/2,\pi\}$. Similarly, the red trajectories on the Bloch spheres correspond to the time evolution when populating the $\ket{X}$ state, i.e., $P_{e,e}$. The dynamics of the $\ket{-}$ population upon excitation from $\ket{G}$ is represented by cyan trajectories. In ((a)-(c)) the target state $\ket{+}$ is being considered $(d/\lambda = 0.01)$, where $\vartheta$ is chosen as \{(a),(b),(c)\}$\mapsto\{0,\pi/2,\pi\}$. In ((d)-(f)) the target state $\ket{e,e}$ $(d/\lambda = 0.1)$ is visualized. Here, we chose $\vartheta$ as \{(d),(e),(f)\}$\mapsto\{0,\pi/2,\pi\}$. Parameters are mentioned in the text.}
\label{fig:bloch2}
\end{figure*}

\subsection{Preparation of Hybrid Collective States with Controlled Optical Phases}
\label{susec:hybrid-states}
As discussed earlier, both the choice of relative phases in the SUPER scheme ($\vartheta_i$) (Eq.~(\ref{Htot})) and in the cavity coupling ($\phi_i$) (Eq.~(\ref{eq:HC})) have an equivalent role, allowing for the creation of hybrid collective states tuning phases of  the SUPER pulses or the cavity coupling. In this section, we discuss the preparation of  collective states, in particular, relying on the relative optical phase $\vartheta$~\cite{lodhal:science:2023, van2025resonant} for the interacting two two-level QEs. The dressed-states picture (DSP) is discussed in detail in Appendix-\ref{apen-dsp}. We consider the Hamiltonian in Eq.~(\ref{eq:HDSPC}) and as discussed in Section.~\ref{susec:Cavity}, the presence of the optical cavity hardly affects the population inversion to a desired collective eigenstate via the SUPER scheme. Especially, the maximum population is independent of the cavity coupling strength. Hence, we neglect the emitter-cavity coupling for a more concise analysis. The system Hamiltonian in the DSP can now be written as
\begin{widetext}
    \begin{equation}
        \hat{H}_{\text{DSP}}(t;\vartheta) = \sum_{j\in\{G,\pm,X\}}E_j\ket{j}\bra{j}\nonumber\\
    -\frac{1}{2}\sum_{j\in\{\pm\}}(\tilde{\Omega}_{\text{S}}(t)(e^{i\vartheta}(\ket{X}\bra{j} \nonumber\\
    + \Xi(j)\ket{j}\bra{G}) \nonumber\\
    + (\Xi(j)\ket{X}\bra{j} + \ket{j}\bra{G})) + \text{~h.c.~})~,
     \label{eq:HDSP}
    \end{equation}
\end{widetext}
where $\tilde{\Omega}_{\text{S}}(t) = \tilde{\Omega}_{\text{S}}^1(t) + \tilde{\Omega}_{\text{S}}^2(t)\exp{i(\Delta_1 - \Delta_2)t}$, $\Xi(\pm) = \pm 1$ and $\hat{H}_{\text{DSP}}(t;\vartheta)$ is essentially a time $t$ and phase $\vartheta$ dependent quantity. Considering the orthonormal dressed-states basis: $\mathcal{B}_{\text{DSP}} = \{\ket{G},\ket{\pm},\ket{X}\}$, $\hat{H}_{\text{DSP}}(t;\vartheta)$ can be written in matrix representation as
\begin{widetext}
\begin{align}
    \hat{H}_{\text{DSP}}(t;\vartheta) = 
    \begin{pmatrix}
        E_X & -\frac{\tilde{\Omega}_{\text{S}}(t)}{2}(e^{i\vartheta} + 1) & -\frac{\tilde{\Omega}_{\text{S}}(t)}{2}(e^{i\vartheta} - 1)  & 0 \\
        -\frac{\tilde{\Omega}^*_{\text{S}}(t)}{2}(e^{-i\vartheta} + 1) & E_+ & 0 & -\frac{\tilde{\Omega}_{\text{S}}(t)}{2}(e^{i\vartheta} + 1) \\
        -\frac{\tilde{\Omega}^*_{\text{S}}(t)}{2}(e^{-i\vartheta} - 1) & 0 & E_- & -\frac{\tilde{\Omega}_{\text{S}}(t)}{2}(-e^{i\vartheta} + 1) \\
        0 & -\frac{\tilde{\Omega}^*_{\text{S}}(t)}{2}(e^{-i\vartheta} + 1) & -\frac{\tilde{\Omega}^*_{\text{S}}(t)}{2}(-e^{-i\vartheta} + 1) & E_G
    \end{pmatrix}.
    \label{eq:HDSP:mat}
\end{align}
\end{widetext}

The above Eq.(\ref{eq:HDSP:mat}) provides further insights about the system, in particular, the interaction between the dressed states and the SUPER pulses. One finds some non-zero off-diagonal terms depending on $\vartheta$, where $\vartheta\in(-\pi,\pi]$. Since these terms quantify the degree of mixture between the $\ket{+}$ and $\ket{-}$ channel, we refer to said terms as \emph{mixture weights}. In Table-\ref{tab2} below, we explicitly write down the mixture weights for certain choices of $\vartheta$.

\begin{table}[h]
\caption{\label{tab2}
Mixture weights of $(\hat{H}_{\text{DSP}}(t;\vartheta))_{i,j} = h_{i,j}$ for different $\vartheta$, where $\{i,j\}\in\mathcal{B}_{\text{DSP}}$.}
\begin{ruledtabular}
\begin{tabular}{ccccc}
$\vartheta$ & $h_{X,+}$ & $h_{X,-}$ &  $h_{+,G}$ & $h_{-,G}$ \\
\midrule
$0$ & $-\tilde{\Omega}_{\text{S}}(t)$ & $0$ & $-\tilde{\Omega}_{\text{S}}(t)$  & $0$  \\
$\frac{\pi}{2}$ & $-\frac{\tilde{\Omega}_{\text{S}}(t)}{2}(i + 1)$ & $-\frac{\tilde{\Omega}_{\text{S}}(t)}{2}(i - 1)$ & $-\frac{\tilde{\Omega}_{\text{S}}(t)}{2}(i + 1)$ & $-\frac{\tilde{\Omega}_{\text{S}}(t)}{2}(-i + 1)$ \\
$\pi$ & $0$ & $\tilde{\Omega}_{\text{S}}(t)$ & $0$ & $-\tilde{\Omega}_{\text{S}}(t)$ \\
\end{tabular}
\end{ruledtabular}
\end{table}

The dynamics of $\hat{H}_{\text{DSP}}(t;\vartheta)$ (Eq.~(\ref{eq:HDSP:mat})) are governed by the master equation (Eq.~(\ref{master}))
\begin{align}
    \partial_t\hat{\rho}_{\text{DSP}}(t;\vartheta) = -i[\hat{H}_{\text{DSP}}(t;\vartheta),~ &\hat{\rho}_{\text{DSP}}(t;\vartheta)]\nonumber\\
    +& \hat{\mathcal{L}}_{\text{Coll}}[\hat{\rho}_{\text{DSP}}(t;\vartheta)]~,
\end{align} 
where we have neglected the cavity. Since $\hat{\rho}_{\text{DSP}}(t;\vartheta)$ acts on the DSP Hilbert space $\mathcal{H}_{\text{DSP}}\cong\mathbb{C}^2\otimes\mathbb{C}^2$, a Bloch sphere description \cite{bloch:PRA:1946} of the problem is not possible. However, we only want to visualize the energy distribution in the system indirectly, therefore, a subspace-projection procedure using the density matrix $\hat{\rho}$ is sufficient. We rewrite the density matrix as
\begin{align}
    \hat{\rho} = \sum_{i,j\in\{G,\pm,X\}}\rho_{i,j}\ket{i}\bra{j}~,
\end{align}
where $\hat{\rho}=\hat{\rho}_{\text{DSP}}(t;\vartheta)$ and $\rho_{i,j}\coloneqq \bra{i}\hat{\rho}\ket{j}$. We continue by considering only the two-level system: $\ket{G}$ \& $\ket{j}$ with $\ket{j}\in\mathcal{B}_{\text{DSP}}\setminus\{\ket{G}\}$ and introduce the projectors $\hat{\Pi}_j = \ket{G}\bra{G} + \ket{j}\bra{j}$. We obtain the sub-density matrix 
\begin{align}
    \hat{\rho}^{j}_{\text{sub}} = \hat{\Pi}_j\hat{\rho}\hat{\Pi}_j = \begin{pmatrix}
        \rho_{j,j} & \rho_{j,G}\\
        \rho_{G,j} & \rho_{G,G} 
    \end{pmatrix}.
\end{align}

We use $\hat{\rho}_{\text{sub}}$ for a Bloch-like interpretation of the dynamics. We do not normalize $\hat{\rho}_{\text{sub}}$, which implies the shortening of the Bloch-like vector (from now on referred to as Bloch vector) is not only caused by decoherence, but also by population reduction of $\ket{j}$. We define the Bloch vector as
\begin{align}
    \textbf{a}_j = \begin{pmatrix}
        2\Re[(\hat{\rho}^{j}_{\text{sub}})_{G,j}]\\
        2\Im[(\hat{\rho}^{j}_{\text{sub}})_{G,j}] \\
        (\hat{\rho}^{j}_{\text{sub}})_{j,j} - (\hat{\rho}^{j}_{\text{sub}})_{G,G}
    \end{pmatrix}.
\end{align}

We apply $\textbf{a}_j$ for the findings in Fig.~\ref{fig3}~(a), i.e., for addressing $\ket{+}$ we choose the optimized SUPER pulse areas $\alpha_1 = 68.25\pi$ and $\alpha_2 = 59.05\pi$. Utilizing the optimization procedure, we can find $\alpha_1 = 51.74\pi$ and $\alpha_2 = 70.48\pi$ to target $\ket{X} = \ket{e,e}$ efficiently (Fig.~\ref{fig2}(e)-(h)). We determine the $\textbf{a}_j$ dynamics here as well. In both cases, we consider $\vartheta\in\{0,\pi/2,\pi\}$. The Bloch vector trajectories are shown in Fig.~\ref{fig:bloch2}. Fig.~\ref{fig:bloch2}((a)-(c)) demonstrate that, when targeting the $\ket{+}$ state, the choice of $\vartheta$ is crucial. A high population inversion can be observed for $\vartheta = 0$ (in Fig.~\ref{fig:bloch2}(a)) with some final oscillations around the north pole, i.e.\ around $\ket{+}$. Red trajectories are present as well, however, note that these contributions never reach a stable vector combination in the vicinity of the $\ket{X}$ north pole. In this case, no cyan-colored $\ket{-}$ contributions are present, that is, as discussed earlier, the `dark' channel can not be populated. The scenario changes when $\vartheta = \pi/2$ (Fig.~\ref{fig:bloch2}(b)) and the contributions of the `dark' channel become dominant when $\vartheta = \pi$ (Fig.~\ref{fig:bloch2}(c)). Hence, in Fig.~\ref{fig:bloch2}(c) the blue trajectories disappear, leaving only red and cyan. However, the collective frequency shift $\Omega_{12}$ is large for $d/\lambda = 0.01$ (Fig.~\ref{schematic}(b)). Therefore, due to the large energy gap in the dressed levels (Fig.~\ref{schematic}(c)), the SUPER parameter sets are not adjusted to ensure high-population inversion for the energy shift $E_+ = -\Delta_1 + \Omega_{12} \to E_- = - \Delta_1 - \Omega_{12}$. Thus, a large final population $P_+$ in $\ket{+}$ does not imply a large population $P_-$ in $\ket{-}$. This is consistent with the above discussions considering the different choices for populating $\ket{\pm}$ and $\ket{X}$. The opposite is the case for $\ket{-}$. However, Fig.~\ref{fig:bloch2}~((d)-(f)) show that the above argument is not valid for populating $\ket{X}$ at $d/\lambda = 0.1$. The choice of $\vartheta$ decides over the contribution of each channel only, But, since the final energy $E_X = -2\Delta_1$ never changes and $E_+ \approx E_-\approx -\Delta_1$, the end result is always the same final population $P_{e,e} = P_X$. In Appendix-\ref{apen-phase} we provide further details on the influence of $\vartheta$ on the populations of the $\ket{\pm}, \ket{X}$ states.

In the following sections, we will consider the cavity interface once again to calculate a few observables for easier photonic measurements in reality.

\begin{figure}[b]
\centering
\includegraphics[width=0.9\linewidth]{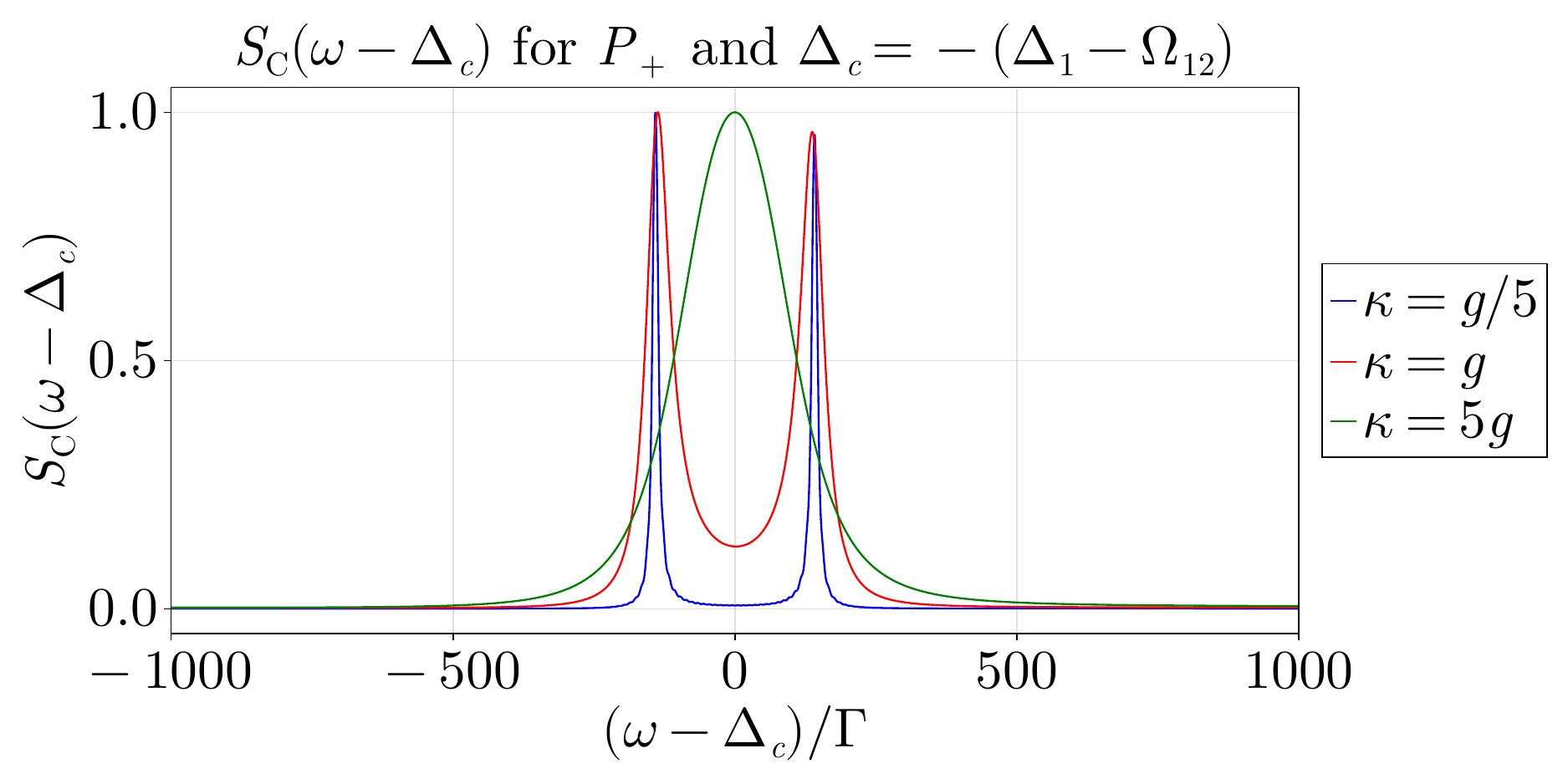}
\caption{Normalized cavity emission spectrum for an efficiently prepared `bright' state.  The interacting QE-cavity system is prepared in the $\ket{+}$ state with SUPER parameters $(\alpha_1 = 68.25\pi, \alpha_2=59.05\pi, \Gamma\tau=0.0)$, but in contrast to Fig.~\ref{fig2}(a)-(d), we choose $\Delta_c = -(\Delta_1 - \Omega_{12})$, allowing the observation of interesting system dynamics. The spectrum calculation is performed until a time when the system has fully relaxed into the ground state, i.e., $\Gamma t\in[0,\Gamma (T-t_0) = 0.6]$. The cavity decay rates are $\kappa\in g\cdot\{1/5,1,5\}$ with $g = 100\Gamma$.}
\label{speccP}
\end{figure}
\section{Cavity Emission Spectrum}
\label{spectra}
The SUPER excitation process is time-dependent. The cavity emission spectrum~\cite{mirza:single:2014, Schumacher:PRR:2024} can be calculated using
\begin{align}
    S_{\text{C}}(\omega) = \lim_{T\to\infty}\int_{0}^{T}\text{d}t\int_{0}^{T-t}\text{d}\tau_{t}~ e^{-i\omega\tau_{t}}\langle \hat{a}^{\dagger}(t)\hat{a}(t+\tau_{t})\rangle~,
    \label{eq:spec:ph}
\end{align}
where the $t$ in $\tau_{t}$ reminds us that $\tau_{t}\in[0, T-t]$ with the upper limit $T$, is chosen such that the system has fully decayed back to the ground state. We consider the case when the superradiant state ($\ket{+}$) for $d = 0.01\lambda$ is targeted (as also in Fig.~\ref{fig2}(b)). In Eq.~(\ref{eq:spec:ph}) we have already substituted $T \to T' = T - t_0 = 0.6~\Gamma$, where $t_0$ is the initial time of the SUPER mechanism to avoid numerical  artifacts. Fig.~\ref{speccP} displays the  normalized spectrum $S_{\text{C}}(\omega) \to (S_{\text{C}}(\omega))'= S_{\text{C}}(\omega) /\max_{\omega}{S_{\text{C}}}(\omega)$ when efficiently preparing $\ket{+}$ from a strong to weak coupling, i.e., $\kappa \in g\cdot\{1/5, 1, 5\}$ via the SUPER scheme. The cavity mode is resonant to the target state $\ket{+}$, i.e.,  $\Delta_c = -(\Delta_1 - \Omega_{12})$ and couples  to the $\ket{+}$ branch only $(\phi_1 = 0 = \phi_2)$. In case of $\kappa = g/5$, Fig.~\ref{speccP} shows two characteristic peaks at shifted frequencies $(\omega - \Delta_c) = \pm 141.11\Gamma$. Increasing $\kappa$ to $\kappa = g$ shifts the peaks slightly towards zero, providing the characteristic frequencies $(\omega - \Delta_c) = \pm 136.91\Gamma$. In the case of a lossy cavity, i.e., $\kappa = 5g$, we obtain on peak at $(\omega - \Delta_c) \approx 0.0\Gamma$. 

One can expect similar features in the emission spectrum when efficiently preparing the dark $\ket{-}$ state, with suitable pulse parameters (as used in Fig.~\ref{fig3}(c)-(d)) and cavity frequency $\Delta_c = -(\Delta_1 + \Omega_{12})$, however, as the subradiant state decays very slowly (as shown in Fig.~\ref{fig3}(d)), $T$ needs to be chosen very large. Therefore, Eq.(\ref{eq:spec:ph}) is numerically expensive and its demonstration is beyond our scope here. When considering hybrid collective states, one could expect similar features. Yet, the choice of the target state will influence the choice of phases $(\vartheta,\phi_i)$ as well as the choice of $\Delta_c$ significantly, which are crucial contributors in the optimization process for near-unity inversion with the SUPER scheme.

Since the emission spectrum with SUPER pulses essentially provides the signature of the cavity mode frequencies, as the targeted levels are experiencing AC-Stark shifts, the feature of Fig.~\ref{speccP} will be similar for other choices of emitter separations $d$ if the cavity mode is close to resonance with the dressed state transition frequency (as in this case) for a suitable choice of pulse parameters. This may also apply to different dipole orientations of QEs. It seems that $S_{\text{C}}(\omega)$ is an energy-level insensitive observable if SUPER pulses are targeting the excitation.

\section{Photon Statistics}
\label{photon}
To understand the photon statistics of the emitted light from the cavity, we compute the photon-photon correlation in terms of the second-order correlation function for the cavity mode $g^{(2)}_{\mathrm{C}}(\tau_f,\tau_t)$, immediately after the preparation of the collective states via the SUPER excitation scheme. Here, $\Gamma\tau_f = 0.02$ indicates the end of the pulses, and the pulses are centered around $\Gamma t=0$. The two-time second-order correlation function has the following form 
\begin{align}
g^{(2)}_C(\tau_f, \tau_{t}) = \frac{\langle \hat{a}^{\dagger}(\tau_f)\hat{a}^{\dagger}(\tau_f+\tau_{t})\hat{a}(\tau_f+\tau_{t})\hat{a}(\tau_f)\rangle}{\langle \hat{a}^{\dagger}(\tau_f)\hat{a}(\tau_f)\rangle\langle\hat{a}^{\dagger}(\tau_f+\tau_{t})\hat{a}(\tau_f+\tau_{t})\rangle}~.
\label{g2tau}
\end{align}

\begin{figure}[h]
\centering
\includegraphics[width=\linewidth]{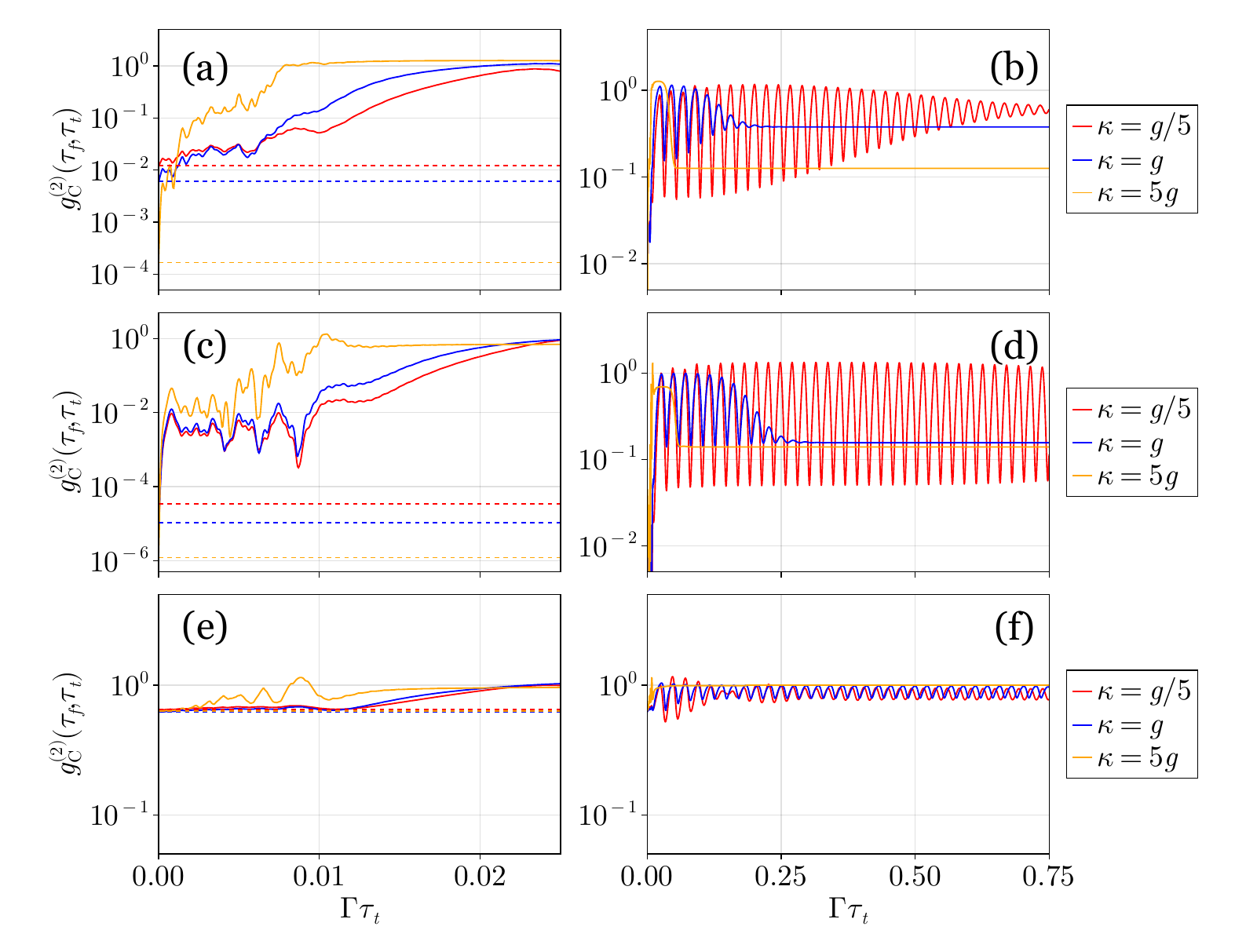}
\caption{Intensity correlation function $g_{\text{C}}^{(2)}(\tau_f,\tau_t)$ for $\tau_f = 0.02~\Gamma$ and with a variable time delay $\tau_t$, after the preparation of the different target states via the SUPER scheme. Different rows in the plots above correspond to signatures after the selective preparation of different target states. The left column, ((a), (c), (d)) displays the short- and the right column, ((b), (d), (f)) shows the long-term variations. In particular, $g_{\text{C}}^{(2)}(\tau_f,\tau_t)$ with time delay $\tau_t$ are visualized when populating the $\ket{+}$ state ((a), (b)), $\ket{-}$ state ((c), (d)) and $\ket{X} = \ket{e,e}$ state ((e), (f)). For (a)-(d), the QEs are at deep-subwavelength separation ($d = 0.01\lambda$) and for (e)-(f)  the separation is $d = 0.1 \lambda$. The pulse parameters are mentioned in the text. The horizontal dashed lines (red, blue and orange) in (a), (c), (e) refer to $g_{\text{C}}^{(2)}(\tau_f,0)$ for the respective coupling regimes (strong, intermediate and weak), with their values indicated in Table-~\ref{g20table}.}
\label{g2tau-fig}
\end{figure}

It measures the probability of detecting a second photon at time $\tau_t$, after the first photon is detected at time $\tau_f$. For anti-bunched non-classical light, one expects $0\leq g^{(2)}_C(\tau_f,\tau_t)<1$. In Fig.~\ref{g2tau-fig}, we plot the variation of $g^{(2)}_C(\tau_f,\tau_t)$ over the scaled time $\Gamma \tau_t$ after the selective preparation of the $\ket{+}$ state with $\Delta_c = -(\Delta_1 - \Omega_{12})$ (short-time (a) and long-time (b)), the $\ket{-}$ state with $\Delta_c = -(\Delta_1 + \Omega_{12})$ (short-time (c) and long-time (d)) and the $\ket{X} = \ket{e,e}$ state with $\Delta_c = -\Delta_1$ (short-time (e) and long-time (f)) from a strong to weak cavity coupling regime, i.e., $\kappa \in g\cdot \{1/5, 1, 5\}$. The corresponding optimized pulse areas together with the time shift are: $(\alpha_1, \alpha_2, \Gamma\tau)$ = $(68.25\pi,59.05\pi, 0.0)$ for target state $\ket{+}$, $(20\pi,40\pi,0.004)$ for state $\ket{-}$ and $(51.74\pi,70.48\pi, 0.0)$ for state $\ket{X}$, respectively. In addition, for (a)-(b) and (e)-(f) $\vartheta = \phi_1 = \phi_2 = 0$, (c)-(d) requires $\vartheta = \pi$, $\phi_1 = 0$ and $\phi_2 = \pi$. In Table-\ref{g20table}, we explicitly note the numerical values of $g^{(2)}_C(\tau_f,0)$ for three different coupling regimes and for the respective target states.

\begin{table}[h]
\caption{\label{tab3} Exact values of $g_{\text{C}}^{(2)}(\tau_f,0)$, i.e., immediately after the preparation of target states $\ket{j}$ ($j\in\{\pm,X\}$) via SUPER excitation scheme at different system-cavity coupling regimes $\kappa\in\{g/5,g,5g\}$.}
\begin{ruledtabular}
\begin{tabular}{cccc}
$\ket{j}$ & $\kappa = g/5$ & $\kappa = g$ &  $\kappa = 5g$ \\
\hline
$\ket{+}$ & $1.20\cdot10^{-2}$ & $6.02\cdot10^{-3}$ & $1.68\cdot10^{-4}$ \\
$\ket{-}$ & $3.49\cdot10^{-5}$ & $1.05\cdot10^{-5}$ & $1.22\cdot10^{-6}$ \\
$\ket{X}$ & $0.65\cdot10^{0}$ & $0.62\cdot10^{0}$ & $0.63\cdot10^{0}$ \\
\end{tabular}
\end{ruledtabular}
\label{g20table}
\end{table}

In Fig.~\ref{g2tau-fig}(a),(c),(e) we observe $g^{(2)}_C(\tau_f, 0) < 1$, implying a single-photon signature, similar to the case of the SUPER excitation scheme addressing single quantum dots~\cite{doris:PRXQ:2021,Schumacher:PRR:2024}. The excitation process by SUPER induces an AC-Stark shift for the target states, therefore, the cavity mode remains off-resonant during the SUPER excitation process. As a result, the system exhibits single-photon emission at $\Gamma\tau_t = 0$. At long times, we find the existence of antibunched light for all three coupling regimes (Fig.~\ref{g2tau-fig}). In Fig.~\ref{g2tau-fig}(b) and in Fig.~\ref{g2tau-fig}(d), similar antibunched photon emission can be observed. For $\kappa=g/5$ (red lines), strong oscillations occur, which imply system-photon finite interaction within the cavity. Increasing $\kappa$ (blue/orange lines) seems to result in damped oscillations, where the system-photon interaction experiences a large reduction. In Fig.~\ref{g2tau-fig}(e), we can find oscillations of $g_{\text{C}}^{(2)}(\tau_f,\tau_t)$ around unity, implying small non-classical contributions at finite $\tau_t$ for $\kappa\in \{g/5,g\}$. For $\kappa = 5g$, we observe $g_{\text{C}}^{(2)}(\tau_f,\tau_t)\approx 1$. 

In Fig.~\ref{fig2}(b) (red curve for $d = 0.01 \lambda$), the maximum population is robust to the change of cavity coupling parameter; however, note that the dissipative temporal behavior depends on the choice of $(\kappa, \Delta_c)$. Larger values of $\kappa$ and off-detuned $\Delta_c$ from the target level (as in Fig.~\ref{fig2}(b)) result in a smooth population decay, in contrast to small $\kappa$ and resonant $\Delta_c$ to the target level (more strong coupling to the environment), which yield oscillatory decay). A consequence of off-resonant $\Delta_c$ is that the cavity mode has a very small number of photons (red line in Fig.~\ref{fig2}(d)) for all coupling regimes. In Table-\ref{g20table}, we find the lowest $g^{(2)}_C(\tau_f, 0)$ at $\kappa = g$ (1.68$\times~10^{-4}$) when $\ket{+}$ is just prepared. This could indicate an approach in the photon domain towards ideal single-photon-states ($g^{(2)}_C(\tau_f, 0)$ = 0) or even quantum non-Gaussian states~\cite{filip:prl:2011,jezek:prl:2011}, i.e., conceptually, those subsets of non-classical states lie deep below ($g^{(2)}_C(\tau_f, 0)\ll 1$) the conventional nonclassical limit~\cite{ivo:prl:2014}. In the strong-coupling regime, $g^{(2)}_C(\tau_f, 0)$ is still low $\sim 10^{-2}$, which indicates that the targeted level $\ket{+}$ is reasonably well decoupled from the environment. However, at weak coupling $\kappa = 5g$, we find $g^{(2)}_C(\tau_f, 0) = 1.68\times~10^{-4} \ll 1$, indicating that environmental decoupling of the target state is much more prominent (as expected). For populating the $\ket{-}$ state, similar features can be observed (see Fig.~\ref{g2tau-fig}(c)-(d)), as this case is also governed by the same fundamental physical principles. At $\kappa = 5g$ the lowest $g^{(2)}_C(\tau_f, 0) = 1.22 \times 10^{-6}$, even lower than for the previous case. Note, $\ket{-}$ is already substantially decoupled from the environment due to its sub-radiant nature. When prepared efficiently, under the illumination by SUPER, the pulses induce AC-Stark shifts, more decoupling from the environment and prohibiting re-excitation results in $g^{(2)}_C(\tau_f, 0) = 3.49 \times 10^{-5} \ll 1$ (see Table-\ref{tab3}), even in the strong system-cavity coupling regime (much lower than the case for state $\ket{+}$). In short, the second-order correlation function provides the same insights as the population dynamics  (Fig.~\ref{fig2}) demonstrated earlier. As discussed in Sec.~\ref{susec:DEC}, other environmental triggers, such as position fluctuations and energy imperfections (upto a reasonable extent), are expected to reflect similar features in $g^{(2)}_C(\tau_f, 0)$ for a deep-subwavelength arrangement.

At $\tau_t = 0$, we obtain $g^{(2)}_C(\tau_f,0) < 1$ for the case of targeting $\ket{X}$ (see Table-\ref{g20table}, the exact values and features are the same as discussed above), showing initially anti-bunched light. In Fig.~\ref{fig2}(h), we obtain a finite population in the cavity (blue curve, largest value $\sim$ 1.5), hence, the cavity is in this case coupled resonantly to the dipole-system. We obtain a small $g^{(2)}_C(\tau_f,0)$, but not as small as for earlier cases. In Fig.~\ref{g2tau-fig}(f), we encounter a numerical instability at longer times (in particular, shortly after $\Gamma \tau_t = 0.1$ due to low mode photon numbers), which we counteracted by scaling the numerator and denominator of $g^{(2)}_C(\tau_f,\tau_t)$ appropriately on top of the simulated outcomes and obtain $g^{(2)}_C(\tau_f,\tau_t) \sim 1$, at long time as displayed in Fig.~\ref{g2tau-fig}(f).

\section{Possibilities for Experimental Realization}

We now consider a few possible experimental strategies to realize the theoretical proposal described here. Several solid-state quantum emitter platforms may be considered for this purpose. Epitaxial semiconductor quantum dots (QDs) offer linewidths (best values reported around 700 MHz), high quantum efficiency, and stable emission characteristics. Typical size distributions of these systems are around 10-70 nm in diameter and 1-10 nm in height. The widely used Stranski-Krastanov growth technique and Droplet epitaxy techniques, however, result in randomly grown QDs~\cite{Gurioli2019}. One could utilize a densely grown QD ensemble and perform statistical analysis of QD distributions to identify pairs that are closely located. Yet this is a time-consuming method and not a scalable one. Site-controlled QDs can be realized through vapor–liquid–solid (VLS) epitaxy, where nanowire structures are grown in a bottom-up process at predetermined locations, in which QDs are embedded during the growth process. This configuration aligns the QD along the optical axis of the nanowire waveguide, and also facilitates vertical stacking of multiple QDs at separations as low as a few nm~\cite{Dalacu2021, laferriere2020multiplexed, laferriere2022unity}. Another approach of interest may be the pick-and-place method, which enables the transfer of a pre-characterized QD from one surface location to another with sub-micrometer accuracy~\cite{Schell2011}. Despite its time-consuming nature, arranging high-quality QDs in desired patterns is feasible. Although colloidal QDs offer unique advantages such as solution processability, size tunability and high-level scalability, their linewidths and coherence properties are potential concerns for our applications. Yet novel materials such as lead halide perovskite nanocrystals \cite{Shamsi2019} may offer potential alternatives here, with reported linewidths at cryogenic temperatures as low as a few meV~\cite{Fu2017}. To arrange such solution-processed QDs in desired patterns with unprecedented accuracy, at separations as low as a few nm, DNA origami may be employed~\cite{Adamczyk2022,Gopinath2021}. With this method, one can expect precise tuning of dipole-dipole interactions to enhance or suppress collective effects. Two-dimensional materials, particularly transition metal dichalcogenides like MoS$_2$, WS$_2$, and WSe$_2$, are another promising platform to implement our proposal, due to their unique optoelectronic properties~\cite{Haider2021,Kumlin2025}. Quantum emitters in these systems are created via localized strain engineering, either using arrays of nanopillars or nanoholes in the underlying substrate ~\cite{Chakraborty2019}. By using electron beam lithography to define the substrate topography, emitter separations can be controlled up to a few nm. Synthetic chromophore arrangements~\cite{Anderson:nc:2022} may also be a platform to realize extremely small inter-molecule separation. Lastly, organic molecules are also possible material platforms to engineer collective effects~\cite{Lange2024,lounis:nc:2022}. They offer much narrower linewidths (around 40 MHz), and are smaller in size, providing significant advantages compared to all the platforms discussed so far. The most popular organic molecules for quantum optics research have been embedded in crystalline hosts such as anthracene or para-terphenyl crystals. While challenges remain, the progress in nanofabrication and engineering suggests that achieving controlled positioning at subwavelength separations is a promising possibility in the foreseeable future. Concurrently, embedding quantum emitters in photonic cavities such as photonic crystal cavities~\cite{lodhal:science:2023,hallett2024controlling,Nobakht2025,Kim2018}, and compensating the energetic disorder~\cite{lodhal:science:2023} together with tuning mechanisms such as electric~\cite{gerardot2007manipulating,bennett2010electric}, strain~\cite{Rastelli2012,Elshaari2018, trotta:prl:2018}, magnetic field~\cite{Koong2020,Phoenix2022}, temperature ~\cite{Kim2018}, or even laser-induced strain~\cite{grim_scalable_2019}, one can engineer collective effects with better control. 

Coming to the experimental implementation of the proposed scheme, multiple approaches are possible. In Ref.~\cite{karli:nl:2022}, the authors employed a broadband femtosecond laser pulse coupled to a frequency-domain 4$f$ pulse shaper to amplitude-shape two desired frequency bands at desired detunings from the target state. On the other hand, in~\cite{joss:nl:2024}, the authors used two different laser sources tuned to the desired detunings that are temporally synchronized (a necessary condition for the SUPER scheme). In practical terms, both of the schemes have their own advantages and disadvantages. While the pulse-shaping technique needs only a single laser source and thereby reduces the resource overhead, the two-laser technique might be advantageous to circumvent the power limitations due to lossy shaping components such as gratings and spatial light modulators. Furthermore, the pulse-shaping technique intrinsically provides temporally overlapped pulses, while the two-laser technique requires additional temporal synchronization electronics. While the pulse shaping technique might be bulky, it is highly flexible and customizable for carving different pulse widths and detunings. On the other hand, the two-laser method is limited in flexibility, depending on the initial specifications (i.e., tuning range, pulse width, etc.). The pulse-shaping technique typically employs a diffraction grating to spread the colours in space, which then gets mapped to the Fourier plane using a focusing element like a lens or a concave mirror. At the Fourier plane, an active modulating element like a spatial light modulator or digital micromirror device can be mounted, which can selectively turn on and off the pixels, enabling intensity control of individual pulses and their specific frequency. Considering the parameter range described in Table-\ref{tab1}, for a QD system emitting around 780 nm, the two required pulses at $\Delta_1 = -5$ meV and $\Delta_2 = -10$ meV detunings with pulse durations around 6 ps with the specified power can be straightforwardly implemented by tuning a broadband 170 fs laser source tuned to 782 nm (which has a total spectral bandwidth of around 15 nm, i.e., 30 meV range). The required power may be independently adjusted by a proper choice of amplitude mask. The versatility of the pulse shaping technique also allows carving arbitrary amplitude profiles for the individual pulses (i.e., Gaussian, rectangular, etc.). To generate two identical sets (cf Fig.~\ref{schematic}(a)), the simplest method is perhaps to use a Michelson interferometer associated with the pulse shaper. The generated photons can be filtered via narrow band notch filters, etalons, or spatial filtering techniques.

For theoretical demonstration, we have used an optical cavity with system-cavity coupling parameter $g \simeq 65.8~\mu$eV~\cite{Schumacher:nc:2015} and a cavity decay rate, $\kappa$. Such coupling strengths are achievable in principle by using high-quality factor photonic cavities such as photonic crystal waveguides, or ring resonators. The decay rate is defined as  $\kappa\in g\cdot\{1/5, 1, 5\}$ to indicate the strong, intermediate, and weak coupling regimes, respectively. It is worth noting that the choice of $g$ is flexible; perhaps that would indicate for choosing a suitable optical cavity interface in reality. A smaller $g$ would likely have a minimal impact on the optimized pulse parameters for inversion into collective states via SUPER, and the maximum population may not be significantly affected. However, normally a weakly coupled cavity would offer better single-photon signature (See Sec.~\ref{photon} for details), as also noted in single QD-cavity case in Ref.~\cite{Schumacher:PRR:2024}. To measure emission spectrum and photon correlations, one typically uses a high-resolution spectrometer and superconducting nanowire single-photon detector (SNSPD) modules compounded with timing electronics. State-of-the-art spectrometers allow for spectral resolution on the order of 0.01 nm (few 10s of $\mu$eV) and photon counting modules provide time resolution close to 10 ps and efficiencies above 90\%.

\section{Conclusions and Outlook}
In this paper, utilizing a Markovian master equation approach, we have theoretically demonstrated that the selective preparation of the collective states of two deep-subwavelength spaced, dipole-coupled two-level QEs is possible using the SUPER excitation scheme. This unconventional excitation scheme, with optimized pulse parameters, facilitates the deterministic preparation of the excitation into \emph{bright} as well as very \emph{dark} collective states in the presence of a dissipative environment. Our investigation shows that the AC Stark shift, imparted by the ultrashort, off-resonate SUPER pulses, on the target collective states help to decouple the excitation from near-resonant environmental modes. This was also previously observed for single QD cases~\cite{doris:PRXQ:2021, Schumacher:PRR:2024}. This fundamental physical principle is crucial for improving access to collective states (to a reasonable extent) by mitigating environmental decoherence factors, especially when compared to other coherent excitation schemes~\cite{Bracht:OQ:23, Schumacher:PRR:2024}. This, in turn, provides the first opportunity (theoretical) to directly observe the signature radiative properties of superradiance and subradiance, in QDs and molecules. Our results on the preparation of hybrid super- and subradiant states with controlled optical phases~\cite{lodhal:science:2023, van2025resonant} may also provide scope for a non-invasive way to produce states equivalent to mixed collective states with atomic disorders~\cite{yelin:pra:2024} and/or spin-phonon hybrid collective state~\cite{olmos:prl:2025}. The detailed discussion on the possible experimental platforms, in particular with solid-state emitters and molecules, suggests that our theoretical predictions are very likely to be realized.

It is important to note that a reasonably controlled system-reservoir interaction could help to confirm our theoretical estimations. The optical cavity interface first allows us to witness the robustness of collective state preparation with SUPER excitation. It then provides opportunities to intuitively expand our observations to include environmental decoherence originating from static position disorder and inhomogeneous broadening. Our scheme also offers scope to witness a single-photon source~\cite{Lounis:rpp:2005, white:nn:2017}, which would be useful for applications in quantum information processing~\cite{jelena:np:2009, hammerer:rmp:2010}; for instance in quantum network nodes~\cite{reiserer:rmp:2022}, quantum repeaters~\cite{Azuma:rmp:2023}, quantum communication and computing~\cite{weihs:nrp:2023}. Our findings suggest that this robustness on the generation of single-photons from collective states is possible even in the presence of a certain degree of environmental decoherence, potentially making the scheme viable at elevated temperatures, as studied in a single QD-cavity system~\cite{Bracht:OQ:23}. This could be a promising development for creating single-photon sources that utilize collective states of interacting QEs and can operate at room temperature~\cite{lounis:nature:2000}. In the context of recent developments in photonic quantum information technologies~\cite{Heindel:aop:23}, this would to be a pertinent question to address. Future experimental confirmation of the preparation or presence of collective states in biological light-harvesting complexes~\cite{Kohler:revbio:2006, nori:np:2013} using the SUPER excitation scheme could provide more enlightening answers about the validity and applicability of our theoretical interacting dipole model in biological geometries~\cite{pal:njp:2025}. These outcomes could contribute to new horizons in the evolving landscape of bio-inspired energy-efficient configurations~\cite{gauger:2024:spie, cardoner:oe:2022, Holzinger:24, gauger:prl:2025, pal:njp:2025}.

\section{Data Availability}
The simulations in this manuscript were performed using the QuantumOptics.jl~\cite{KRAMER2018}, CollectiveSpins.jl~\cite{collectivespins}  and QuantumToolbox.jl~\cite{QuantumToolbox-jl2025} frameworks in the Julia programming language. The plots were prepared using the Julia framework library~\cite{makie:2021}. The data supporting the findings of this manuscript are openly available~\cite{kerber:zenodo}.

\section*{Acknowledgements}
We thank Gabriela Militani, Charlie Evagora, and René Schwarz for fruitful discussions. This research was funded in whole or in part by the Austrian Science Fund (FWF) 10.55776/ESP682. We also acknowledge the funding from the Austrian Science Fund (FWF) project 10.55776/FG5 (Forschungsgruppe FG 5), 10.55776/COE1 (quantA), and 10.55776/TAI556 (DarkEneT).

\appendix
\section{Dressed-State Picture of Two Interacting Quantum Emitters Inside an Optical Cavity}
\label{apen-dsp}
The bare-basis definition of the Hamiltonian $\hat{H}(t)$ in Eq.~(\ref{eq:Ham1}) is chosen such that the SUPER scheme and the cavity mode can arbitrarily couple to the eigenstates of the dipole-dipole Hamiltonian $\hat{H}_{\text{DD}}$. To gain more insight into the system, such as defining collective super- or subradiance, we introduce the dressed-states picture (see Fig.~\ref{schematic}(c)). We begin by reconsidering Eq.~(\ref{eq:Hdd}), in the rotating frame of $\omega_1$ as following:
\begin{align}
    \hat{H}_{\text{DD}} = -\Delta_1(\hat{\sigma}_1^+\hat{\sigma}_1^- + \hat{\sigma}_2^+\hat{\sigma}_2^-) + \Omega_{12}(\hat{\sigma}_1^+\hat{\sigma}_2^- + \hat{\sigma}_1^-\hat{\sigma}_1^+),\label{eq:Hdd1}
\end{align}
where $\Omega_{12} = \Omega_{21}.$ We can write Eq.~(\ref{eq:Hdd1}) in matrix representation:
\begin{align}
    \hat{H}_{\text{DD}} = \begin{pmatrix}
    -2\Delta_1 & 0 & 0 & 0 \\
    0 & -\Delta_1 & \Omega_{12} & 0 \\
    0 & \Omega_{12} & - \Delta_1 & 0 \\
    0 & 0 & 0 & 0
    \end{pmatrix}~.
    \label{eq:mat}
\end{align}
As mentioned in Sec.~\ref{sec:TD}, the eigenstates are
\begin{align}
\ket{G} =& \ket{g,g}~,\nonumber\\
\ket{\pm} =& \frac{1}{\sqrt{2}}\big(\ket{g,e} \pm \ket{e,g}\big),\\
\ket{X} =& \ket{e,e}~,\nonumber    
\end{align}
which we refer to as dressed states. The corresponding eigenenergies are $E_G = 0$, $E_{\pm} = -\Delta_1 \pm \Omega_{12}$ and $E_X = -2\Delta_1.$ Note that, being a distance and dipole orientation dependent quantity, the coherent coupling $\Omega_{12}$ lifts the energy degeneracy of the first excitation manifold for finite distance separation $d$ between the QEs. We rewrite Eq.~(\ref{eq:Ham1}) in the dressed basis (Eq.~\ref{eq:mat})
\begin{align}
    \hat{\sigma}_1^+ &= \hat{\sigma}^+\otimes\mathds{1}~,\nonumber\\
    &= \ket{e}\bra{g}\otimes \big(\ket{e}\bra{e} + \ket{g}\bra{g}\big)~,\nonumber\\
    &= \ket{e,e}\bra{g,e} + \ket{e,g}\bra{g,g}~,\nonumber\\
    &=\frac{1}{\sqrt{2}}\big(\ket{X}\bra{+} + \ket{X}\bra{-} + \ket{+}\bra{G} - \ket{-}\bra{G}\big)~,\nonumber\\
    \hat{\sigma}_2^+ &= \frac{1}{\sqrt{2}}\big(\ket{X}\bra{+} - \ket{X}\bra{-} + \ket{+}\bra{G} + \ket{-}\bra{G}\big)~,
\end{align}
where $\ket{e,g} = 1/\sqrt{2}(\ket{+} - \ket{-})$ and $\ket{g,e} = 1/\sqrt{2}(\ket{+} + \ket{-})$. Then, the SUPER terms in Eq.~(\ref{eq:Ham1}) $(\hat{\sigma}_1^+e^{i\vartheta} + \hat{\sigma}_2^+)$ can be written as follows:
\begin{align}
    \hat{\sigma}_1^+e^{i\vartheta} &+ \hat{\sigma}_2^+ = \nonumber\\
    &\frac{e^{i\vartheta}}{\sqrt{2}}\big(\ket{X}\bra{+} + \ket{X}\bra{-} + \ket{+}\bra{G} - \ket{-}\bra{G}\big)\nonumber\\
    &+ \frac{1}{\sqrt{2}}\big(\ket{X}\bra{+} - \ket{X}\bra{-} + \ket{+}\bra{G} + \ket{-}\bra{G}\big).
\end{align}
Choosing $\vartheta = 0$ results in
\begin{align}
    \hat{\sigma}_1^+ + \hat{\sigma}_2^+ = \sqrt{2}\big(\ket{X}\bra{+} + \ket{+}\bra{G}\big).
\end{align}
In this case, the SUPER terms in the dressed-states picture can be written as the following Hamiltonian:
\begin{align}
    \hat{\tilde{H}}_{\text{SUPER}} &= -2\frac{\tilde{\Omega}^1_{\text{S}}(t)}{2}\big(\ket{X}\bra{+} + \ket{+}\bra{G} + \text{~h.c.~}\big)\nonumber \\
&-2\frac{\tilde{\Omega}^2_{\text{S}}(t-\tau)}{2}\big(\big(\ket{X}\bra{+} + \ket{+}\bra{G}\big)e^{i(\Delta_1 - \Delta_2)t}\nonumber\\
&+ \text{~h.c.~}\big)~,
\label{eq:SUPERDSP}
\end{align}
where $\tilde{\Omega}^i_{\text{S}}(t) = \Omega^i_{\text{S}}(t)/\sqrt{2}$. For $\vartheta = 0$, $\hat{\tilde{H}}_{\text{SUPER}}$ couples only to the symmetric $\ket{+}$ branch. Similarly, $\vartheta = \pi$ leads to the SUPER pulses only coupling to the anti-symmetric $\ket{-}$ branch. For the cavity Hamiltonian $\hat{H}_{\text{C}}$ (in Eq.~(\ref{eq:HC}) the same logic will remain applicable. For instance, the choice $\phi_1 = 0 = \phi_2$ leads to the mode only coupling to the $\ket{+}$ branch and, in contrast, for $\phi_1 = 0$ and $\phi_2 = \pi/2$ the cavity mode couples equally to branches $\ket{+}$ and $\ket{-}$ as shown below for the interaction part of $\hat{H}_{\text{C}}$
\begin{align}
    g(\hat{\sigma}_1^+ + \hat{\sigma}_2^+e^{i\pi/2})\hat{a}+\text{h.c.}
    &= \frac{1}{\sqrt{2}}(\tilde{g}(\ket{X}\bra{+} + \ket{+}\bra{G}) \nonumber\\
    &+\tilde{g}^*(\ket{X}\bra{-} - \ket{-}\bra{G}))\hat{a}+\text{h.c.}~.
    \label{eq:cavityDSP}
\end{align}
The coupling constant $g \to \tilde{g} = g(1+i)$ is now complex. The Hamiltonian in Eq.~(\ref{eq:Ham1}) in the dressed-states picture (DSP) takes on the general form:
\begin{widetext}
\begin{align}
    \hat{H}_{\text{DSP}}(t) &= \sum_{j\in\{G,\pm,X\}}E_j\ket{j}\bra{j} -\Delta_1\hat{a}^{\dagger}\hat{a} \nonumber\\
    &+\frac{g}{\sqrt{2}}\sum_{j\in\{\pm\}}(\hat{a}(e^{i\phi_1}(\ket{X}\bra{j} + \Xi(j)\ket{j}\bra{G}) + e^{i\phi_2}(\Xi(j)\ket{X}\bra{j} + \ket{j}\bra{G})) + \text{~h.c.~})\nonumber\\
    &-\frac{\tilde{\Omega}_{\text{S}}^1(t)}{2}\sum_{j\in\{\pm\}}(e^{i\vartheta}(\ket{X}\bra{j} + \Xi(j)\ket{j}\bra{G}) + (\Xi(j)\ket{X}\bra{j} + \ket{j}\bra{G}) + \text{~h.c.~})\label{eq:HDSPC}\\
    &-\frac{\tilde{\Omega}_{\text{S}}^2(t-\tau)}{2}\sum_{j\in\{\pm\}}(e^{i(\Delta_1 - \Delta_2)t}(e^{i\vartheta}(\ket{X}\bra{j} + \Xi(j)\ket{j}\bra{G}) + (\Xi(j)\ket{X}\bra{j} + \ket{j}\bra{G})) + \text{~h.c.~})~,\nonumber
\end{align}
\end{widetext}
where $\Xi(\pm) = \pm1$. The collective decay rates $\Gamma_{\pm} = \Gamma\pm\Gamma_{12}$ can be obtained simply by rewriting the Liouvillian superoperator in terms of the dressed states \cite{ostermann:collective:2016}. For $\Gamma_{12}>0$, the $\ket{+}$ branch is the super- and the $\ket{-}$ branch is the subradiant decay channel. The branches switch roles when $\Gamma_{12} < 0$.
\begin{figure}[b]
\centering
\includegraphics[width=\linewidth]{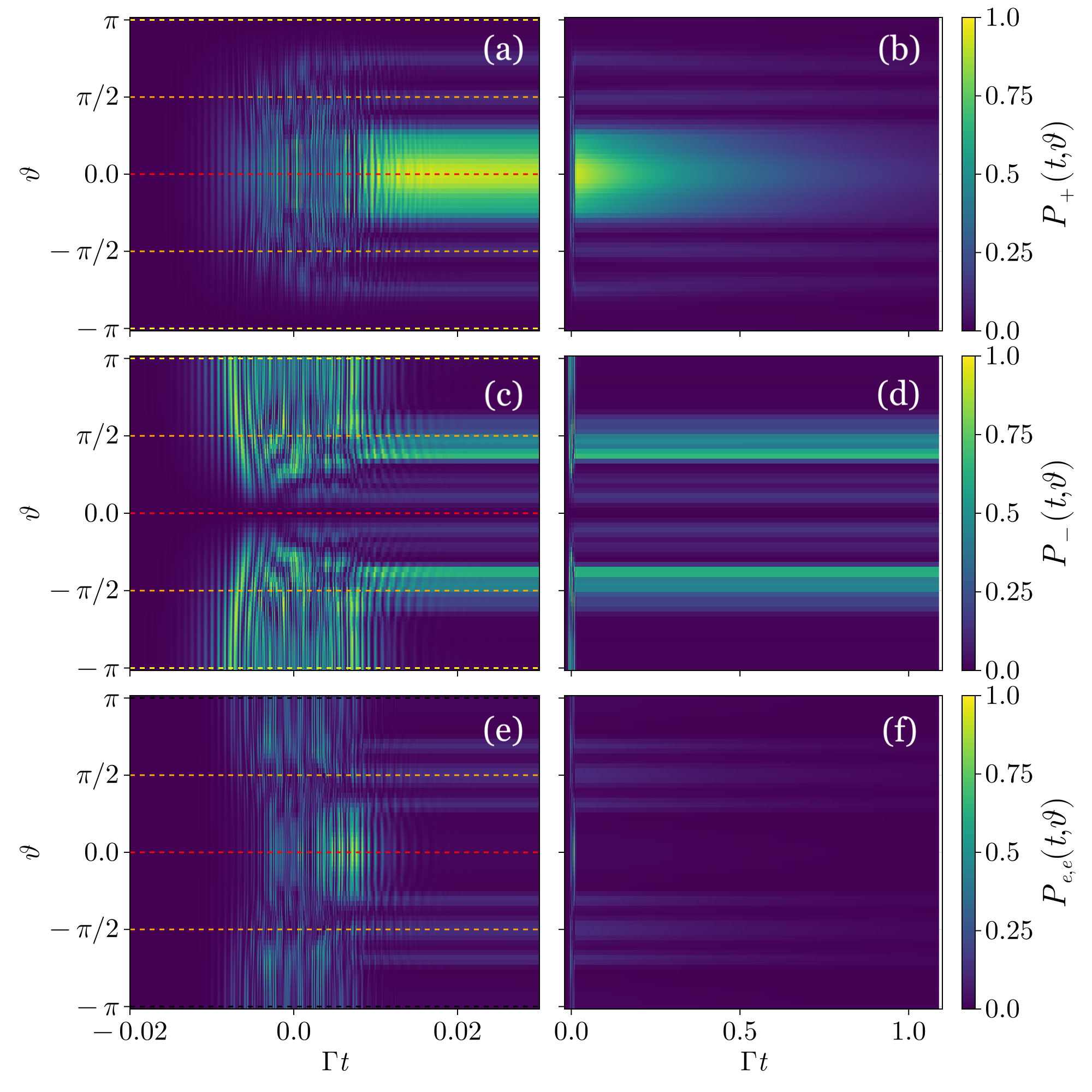}
\caption{The populations $P_{+}(t,\vartheta)$ ((a), (b)), $P_{-}(t,\vartheta)$ ((c), (d)) and $P_{e,e}(t,\vartheta)$ ((e), (f)) with $\vartheta\in(-\pi,\pi]$ are shown when $d = 0.01\lambda$. Red, orange, and yellow/black horizontal dashed lines in (a), (c), (e) correspond to $\vartheta = 0, \pm \pi/2, \pm \pi$, indicating the dynamical trajectories in the Bloch-sphere as shown in Fig.~\ref{fig:bloch2}(a), (b) and (c), respectively.}
\label{fig:phase:plus}
\end{figure}
\begin{figure}[b]
\centering
\includegraphics[width=\linewidth]{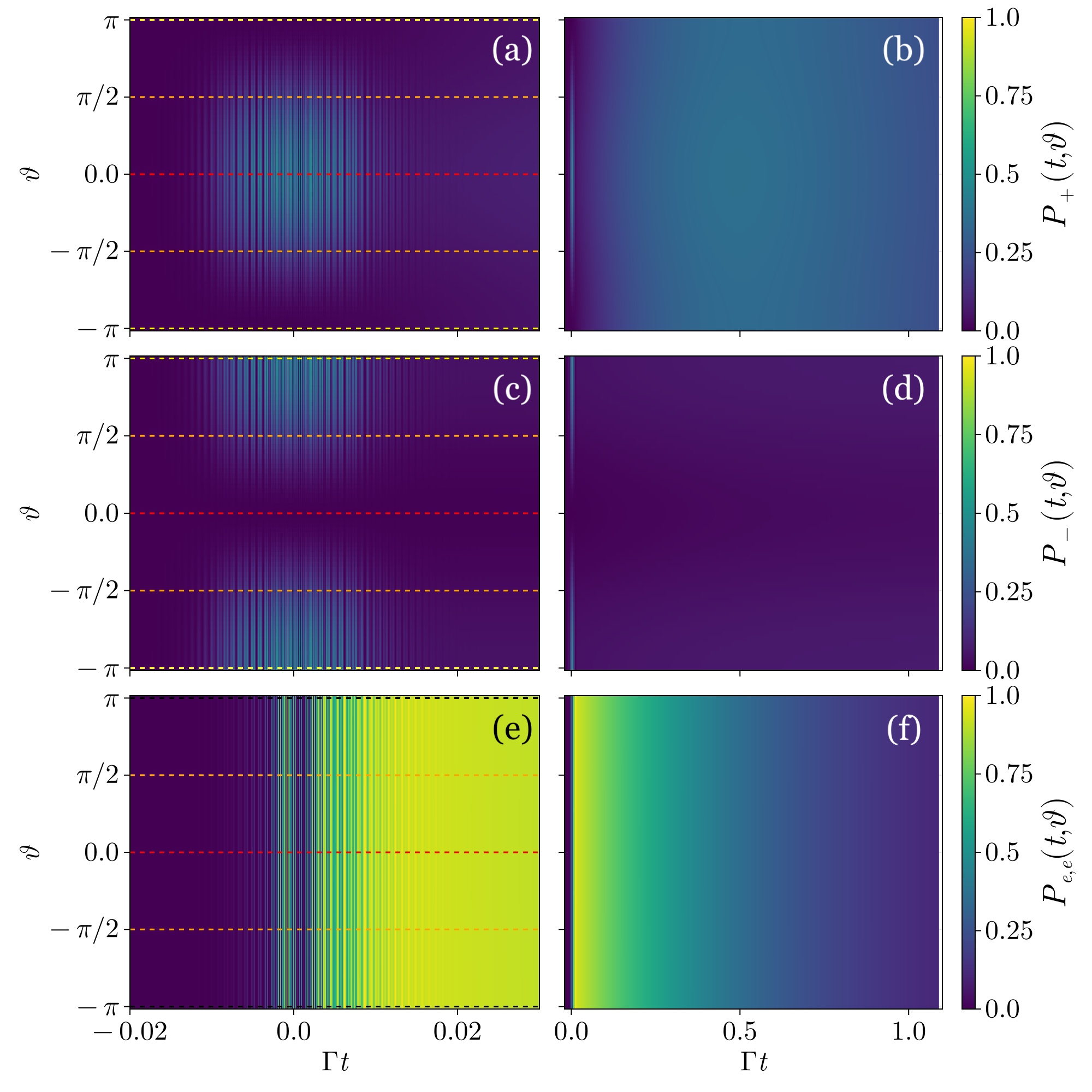}
\caption{The populations $P_{+}(t,\vartheta)$ ((a), (b)), $P_{-}(t,\vartheta)$ ((c), (d)) and $P_{e,e}(t,\vartheta)$ ((e), (f)) with $\vartheta\in(-\pi,\pi]$ are shown for $d = 0.1\lambda$. Red, orange and yellow/black horizontal dashed lines in (a), (c), (e) correspond to $\vartheta = 0, \pm \pi/2, \pm \pi$, representing the dynamical trajectories in the Bloch-sphere as shown in Fig.~\ref{fig:bloch2}(d), (e), and (f), respectively.}
\label{fig:phase:X}
\end{figure}

Considering Eq.~(\ref{eq:SUPERDSP}), we recognize that the two identical SUPER pulses in $\hat{H}(t)$, acting on the two QEs separately, interact in the dressed-states picture in a way such that the beat-like interference of the two Gaussian pulses spreads over the entire $\ket{+}$ branch: $\ket{X}\bra{+} + \ket{+}\bra{G}$. Depending on the optimized parameters discussed in Sec.~\ref{inversion-nocavity}, we can address $\ket{+}$ or $\ket{X}$ directly. For $\vartheta = \pi$, this is true for the $\ket{-}$ branch. A more detailed discussion is given in Appendix-\ref{appen-inf}.


\section{The Influence of Phases in Populating Excited States of Coupled Quantum Emitters}\label{appen-inf}
\label{apen-phase}
As discussed in Sec.~\ref{susec:hybrid-states}, the Bloch representation is very useful to visualize the population dynamics of the system with the SUPER scheme. However, this representation does not provide complete information about the influence of $\vartheta$ on the state populations $P_{\pm}$ and $P_{e,e}$ with the relative phase $\vartheta$. Here, we specifically plot the populations $P_{\pm}(t,\vartheta)$ ((a)-(d)) and $P_{e,e}(t,\vartheta)$ ((e),(f)) in Fig.~\ref{fig:phase:plus} and in Fig.~\ref{fig:phase:X}, respectively, explicitly pointing out the dependence on $\vartheta$ and the temporal evolution. In plots (a),(c),(e), the time evolutions during the SUPER excitation is displayed, whereas (b),(d),(f) present the corresponding long-time evolutions of the system. Fig.~\ref{fig:phase:plus} shows the variations for $d = 0.01\lambda$ and with chosen pulse areas $\alpha_1=68.25\pi$ and $\alpha_2 = 59.05\pi$  ($\ket{+}$ is the target state). The horizontal lines indicate the corresponding Bloch-sphere dynamics that are denoted by red-dashed cut at $\vartheta =0$ (see. Fig.~\ref{fig:bloch2}(a)), orange-dashed cut at $\vartheta = \pm \pi/2$ (for Fig.~\ref{fig:bloch2}(b)) and yellow/black-dashed cut at $\vartheta = \pm \pi$ (in Fig.~\ref{fig:bloch2}(c)). Similarly in Fig.~\ref{fig:phase:X}, where we use $\alpha_1 = 51.74\pi$ and $\alpha_2 = 70.48\pi$ at $d=0.1\lambda$ ($\ket{X}$ is the target state). The corresponding Bloch dynamics are presented in Fig.~\ref{fig:bloch2}((d)-(f)).

Considering the case of the target state $\ket{+}$, Fig.~\ref{fig:phase:plus} (a) clearly shows that the choice of $\vartheta$ is crucial. This is expected, since $\vartheta$ dictates the values of the mixture weights $h_{i,j}$ in Eq.~(\ref{eq:HDSP:mat}). The off-diagonal $h_{i,j}$ mediate the SUPER mechanism for the transition $\ket{j}\to\ket{i}$. The SUPER parameters are optimized for the ideal case of $\vartheta = 0$, i.e. $h_{X,+} = h_{+,G}= -\tilde{\Omega}_{\text{S}}(t) \neq 0$ and $h_{X,-} = h_{-,G}=0$ (Table-\ref{tab2}). The choice $\vartheta\neq 0$ allows the excitation to penetrate into the $\ket{-}$ branch. This reduces the final population significantly. One would expect a small, even no, increase of the final population in $\ket{-}$ for $\vartheta\to\pm\pi$, since the collective fine-structure splitting $\Omega_{12}$ implies $E_{+}\gg E_-$. This can be observed. The final population of $\ket{X}$ remains small in this case. 

Focusing now on the efficient population of $\ket{X}$ in Fig.~\ref{fig:phase:X}(c), an interesting behavior emerges. We find that the choice of $\vartheta$ does not influence the outcome of the final population in $\ket{X}$ at all, even though the mixture weights $h_{i,j}\neq0$ allow contributions of both, the $\ket{+}$ and $\ket{-}$ branch. One might naively expect this to always apply to the $\ket{X}$ state preparation, regardless of the inter-emitter separation $d$, because the final energy $E_X$ remains the same and the SUPER pulses constructively interfere with each other. However, for example, while we can find optimized parameters to efficiently populate $\ket{X}$ with $\vartheta = \pi$ (using the $\ket{-}$ branch) at $d = 0.01\lambda$ (not shown in this manuscript), we observe close-to-zero inversions $\forall\ket{j}\in\{\ket{\pm},\ket{X}\}$ if $0 < |\vartheta|<\pi$. This implies that the SUPER pulse interferences are non-trivial and are greatly influenced by the mixture weights $h_{i,j}$ ($h_{i,j}^*$) as well as $\Omega_{12}$.  In future work, we will specifically focus on how these dependencies influence the final population of the $\ket{X}$ state.

\bibliography{draftbib}

\end{document}